\documentclass[usenatbib]{mnras}

\usepackage[T1]{fontenc}
\usepackage{ae,aecompl}

\usepackage{url}
\usepackage{amsmath}	
\usepackage{graphicx}	

\newcommand{\PRESTO}{\texttt{PRESTO}}

\newcommand{\PSRCHIVE}{\texttt{PSRCHIVE}}

\newcommand{\s}{~{\rm s}}

\newcommand{\dmunit}{pc\,cm$^{-3}$}

\newcommand{\msun}{M$_{\odot}$}

\newcommand{\Mp}{M_{\rm p}}
\newcommand{\Mc}{M_{\rm c}}

\newcommand{\xp}{x_{\rm p}}
\newcommand{\xc}{x_{\rm c}}

\newcommand{\pmunit}{~{\rm mas~yr}^{-1}}

\title[Timing NGC~1851A with the upgraded GMRT]{Upgraded Giant Metrewave Radio Telescope timing of NGC~1851A: a possible millisecond pulsar $-$ neutron star system}
\author[A. Ridolfi et al.]{\parbox{\textwidth}{
A. Ridolfi,$^{1,2}$\thanks{E-mail: ridolfi@mpifr-bonn.mpg.de} P. C. C. Freire$^{1}$, Y. Gupta$^{3}$ and S.M. Ransom$^{4}$
}
\vspace{0.4cm} \\
\parbox{\textwidth}{
$^{1}$Max Planck Institute for Radio Astronomy, Auf dem Hugel 69, D-53121
Bonn, Germany\\
$^{2}$INAF -- Osservatorio Astronomico di Cagliari, Via della Scienza 5, I-09047 Selargius (CA), Italy\\
$^{3}$National Centre for Radio Astrophysics, TIFR, Pune University Campus, Post Bag 3, Pune 411 007, India\\
$^{4}$National Radio Astronomy Observatory, 520 Eclgemont Rcl., Charlottesville, VA, 22903 USA
}
}
\date{Accepted XXX. Received YYY; in original form ZZZ}

\pubyear{2016}

\begin{document}
\label{firstpage}
\pagerange{\pageref{firstpage}--\pageref{lastpage}}
\maketitle

\begin{abstract} In this work we present the results of one year of upgraded Giant Metrewave Radio
Telescope timing measurements of PSR~J0514$-$4002A, a 4.99-ms pulsar in a 18.8-day, eccentric ($e \, =\, 0.89$) orbit with a
massive companion located in the globular cluster NGC~1851. Combining these data with earlier Green Bank Telescope data, we
greatly improve the precision of the rate of advance of periastron, $\dot{\omega} \, = \, 0.0129592(16)\, \deg \, \rm
yr^{-1}$ which, assuming the validity of general relativity, results in a much refined measurement of the total mass of the
binary, $M_{\rm tot} \, = \, 2.4730(6) \, M_{\odot}$. Additionally, we measure the Einstein delay parameter, $\gamma$,
something that has never been done for  any binary system with an orbital period larger than $\sim \,$10 h. The measured
value, $\gamma \, = \, 0.0216(9) \, \rm s$, is by far the largest for any binary pulsar. Furthermore, we measure the proper
motion of the system ($\mu_{\alpha} \, = \, 5.19(22)$ and $\mu_{\delta} = -0.56(25)\rm~mas ~ yr^{-1}$), which is not only
important for analyzing its motion in the cluster, but is also essential for a proper interpretation of $\gamma$, given the
latter parameter's correlation with the variation of the projected semi-major axis. The measurements of $\gamma$ and the
proper motion enable a separation of the system component masses: we obtain a pulsar mass of $M_{p} \, = \,
1.25^{+0.05}_{-0.06} \, M_{\odot}$ and a companion mass of $M_{c} \, = \, 1.22^{+0.06}_{-0.05} \, M_{\odot}$. This raises
the possibility that the companion is also a neutron star. Searches for radio pulsations from the companion have thus far
been unsuccessful, hence we cannot confirm the latter hypothesis. The low mass of this millisecond pulsar - one of the
lowest ever measured for such objects - clearly indicates that the recycling process can be achieved with a relatively small
amount of mass transfer.
\end{abstract}





\section{Introduction}
\label{sec:intro}

The physical conditions found in globular clusters (GCs) show remarkably different characteristics when compared to those in
our Milky Way (e.g., \citealt{Ransom2008,2013IAUS..291..243F}). The density of stars reached near the cores of GCs can
easily exceed by several order of magnitudes the typical values found in the Galactic plane. This translates into an
exceptionally high probability of gravitational interactions between stellar systems, which in turn promotes the formation
(but also the disruption) of two- or many-body bound systems \citep{2014A&A...561A..11V}. As a result, GCs are extremely
prolific hotbeds for the formation of exotic systems that, as exemplified below, are the result of non-standard paths of
binary evolution. Among these exotic systems is a large number of binary millisecond pulsars (e.g.
\citealt{2013IAUS..291..243F})\footnote{For a complete list of all known pulsars in globular clusters, see
\url{https://www.naic.edu/~pfreire/GCpsr.html}.}. \vskip 1 cm As a consequence of their unusual formation paths, the
population of radio pulsars in GCs shows striking differences with respect to that of our Galaxy; consisting almost entirely
of extremely old, recycled pulsars (in fact, the mystery is the presence of some anomalously young pulsars, see e.g.
\citealt{2011ApJ...742...51B}, a likely solution to this problem is the recent disruption of LMXBs, see
\citealt{2014A&A...561A..11V}). Although the total number of globular cluster pulsars accounts for about 5\% of the total
pulsar population, the millisecond pulsars (MSPs, here defined as those having a spin period $P<10$~ms) that are found in
GCs account for about 40\% of the known MSP population\footnote{As of 2018 April, see the \texttt{PSRCAT} pulsar catalog at
\url{http://www.atnf.csiro.au/research/pulsar/psrcat} \citep{Manchester2005}}.

This large population of MSPs include some of the most extreme pulsars and systems known. Among these are extremely recycled
MSPs (e.g., PSR J1748$-$2446ad in Terzan 5; \citealt{Hessels+2006}), extremely compact binaries (e.g., PSR J0024$-$7204R in
47 Tucanae; \citealt{Freire+2017}), extremely energetic pulsars, with very high $\gamma$-ray luminosities (e.g.
PSR~B1820$-$30 in NGC~6624, \citealt{2011Sci...334.1107F}; and PSR~B1821$-$24 in M28, \citealt{2013ApJ...778..106J}) and a
radio pulsar in a ``redback'' system that changes into an accreting X-ray MSP and back to radio in timescales of weeks
\citep{2013Natur.501..517P}.

Among the most exotic systems in GCs are a small group of MSPs in highly eccentric binaries with massive companions: PSR
J1835$-$3259A in NGC 6652 \citep{DeCesar+2015}, PSR~J1807$-$2500B \citep{2012ApJ...745..109L}, and PSR J0514$-$4002A in NGC
1851, \citep{Freire+2004,Freire+2007}. These systems are so unlike anything seen in the Galaxy that they are almost
certainly the result of {\em secondary exchange encounters}, i.e., exchange encounters that happen after the pulsar is
recycled by a lower-mass companion, which is ejected and exchanged by a much more massive compact object. This is only
likely to happen (and has only been observed) in GCs with a large interaction rate {\em per binary} 
\citep{2014A&A...561A..11V}. A confirmation of this is PSR~B2127+11C, a double neutron star system in the core-collapsed
globular cluster M15 \citep{2006ApJ...644L.113J}. Although it superficially resembles a ``normal'' double neutron star like
those found in the Galaxy, it was recognized at an earlier stage that it too must be the result of a secondary exchange
interaction \citep{1991ApJ...374L..41P}. 

These systems are the ultimate example of a non-standard evolutionary path; they suggest that even more exotic systems, like
double MSP and MSP-black hole systems, might be discovered in the future.

\subsection{NGC~1851A}

One of these systems, PSR~J0514$-$4002A, consists of a 4.99-ms pulsar in orbit around a massive companion every 18.8 days in
a very eccentric ($e=0.89$) orbit. It is located in the globular cluster NGC~1851, henceforth we designate this system as
NGC~1851A. The pulsar was discovered at 327 MHz with the Giant Metrewave Radio Telescope (GMRT) located near Khodad, India,
in the context of a small low-frequency survey for pulsars in GCs \citep{Freire+2004}.

Subsequent follow-up observations with the Green Bank Telescope allowed for the derivation of a phase-connected timing
solution \citep{Freire+2007}. As suggested by earlier GMRT interferometric images \citep{Freire+2004}, the pulsar is at
about one core radius from the centre of NGC 1851. By measuring the rate of periastron advance, the authors were able to
derive the total mass of the system ($M_{\rm tot}=2.453(14)$~\msun) and obtain an upper limit on the mass of the pulsar
($\Mp < 1.5$~\msun) and a lower limit on that of the companion ($\Mc > 0.96$~\msun). Although the data were not sufficient
to detect additional post-Keplerian (PK) effects, which would give access to the individual masses of the binary components,
\citet{Freire+2007} envisaged a measurement of the Einstein delay ($\gamma$), and, depending on the system inclination, of
the Shapiro delay, in a not too distant future.

\subsection{Motivation and structure of the paper}

After a hiatus of about one decade, new observations of NGC~1851A were motivated by the major upgrade recently undergone by
the GMRT \citep{Gupta+2017}. In a first stage, issues with the timing stability have been solved, allowing for precise
timing of MSPs. Later, the whole array has been upgraded with new receivers and electronics.  Compared to its original
configuration, the upgraded GMRT (hereafter, uGMRT) delivers up to a factor of three better sensitivity. This is achieved
mainly by means of new wide-band receivers, which provide an almost seamless frequency coverage from $\sim 50$ to $\sim
1450$~MHz, and a modern digital backend system \citep{Reddy+2017} that allows a maximum instantaneous bandwidth of 400 MHz
with real-time coherent de-dispersion. The latter feature is critical to maximize the sensitivity to far, highly dispersed
GCs, when observed at very low frequencies. 

At the same time, detailed simulations suggested that a dense timing campaign over one orbit should be able to determine at
least the relativistic $\gamma$ parameter with enough accuracy to measure the component masses to an uncertainty of about
$0.05 \, M_{\odot}$. Furthermore, a more sparse set of timing measurements spread over one year would enable a precise
measurement of the proper motion of the system. This, as shown below, is of great importance for a proper interpretation of
the measurement of $\gamma$.

Thus, the new capabilities of the uGMRT offered the chance to finally measure the mass of NGC~1851A and its companion with
good precision. Such a measurement is extremely valuable, not only for improving the statistics of MSP mass measurements
(still very small in GCs), but also to investigate the nature of the companion. As we will see below, our measurements
indicate that this companion could well be another neutron star (NS). Since the system very likely formed in an exchange
encounter, the nature of this companion cannot be elucidated by arguments based on stellar evolution.

The remainder of the paper is organized as follows: In section~\ref{sec:observations}, we describe the new uGMRT timing
observations and how the resulting data were reduced. In section~\ref{sec:results}, we present the results of our timing
analysis, with a particular emphasis on the measurements of the PK parameters and their likely kinematic contaminants; this
will be especially relevant for the measurement of the variation of the orbital period ($\dot{P}_{\rm b}$) and $\gamma$. In
section~\ref{sec:masses}, we perform a self-consistent Bayesian analysis of the orbital orientation space to determine the
likely inclination ranges and the masses of the components. Since the companion has a mass that is compatible with it being
a NS, it might also be a pulsar. For this reason, in section~\ref{sec:companion_search} we search for pulsations from the
companion. 
globular cluster NGC~1851.  Finally, in section~\ref{sec:conclusions}, we summarize our findings, elaborate on the nature of
the companion star and discuss some interesting long-term prospects.

\section{Observations and data reduction}
\label{sec:observations}

\begin{figure}
\centering
\includegraphics[width=0.48\columnwidth]{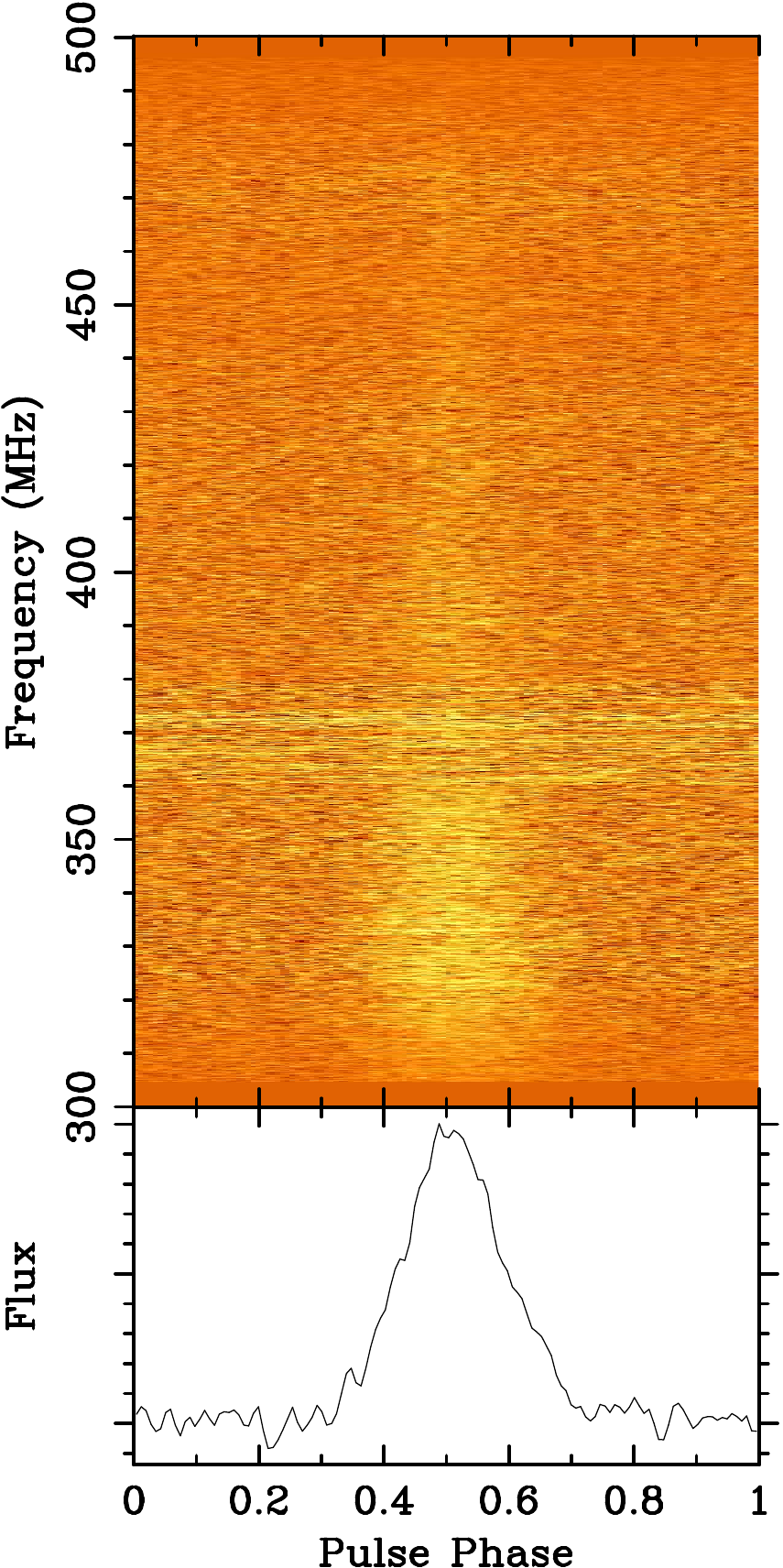}
\includegraphics[width=0.48\columnwidth]{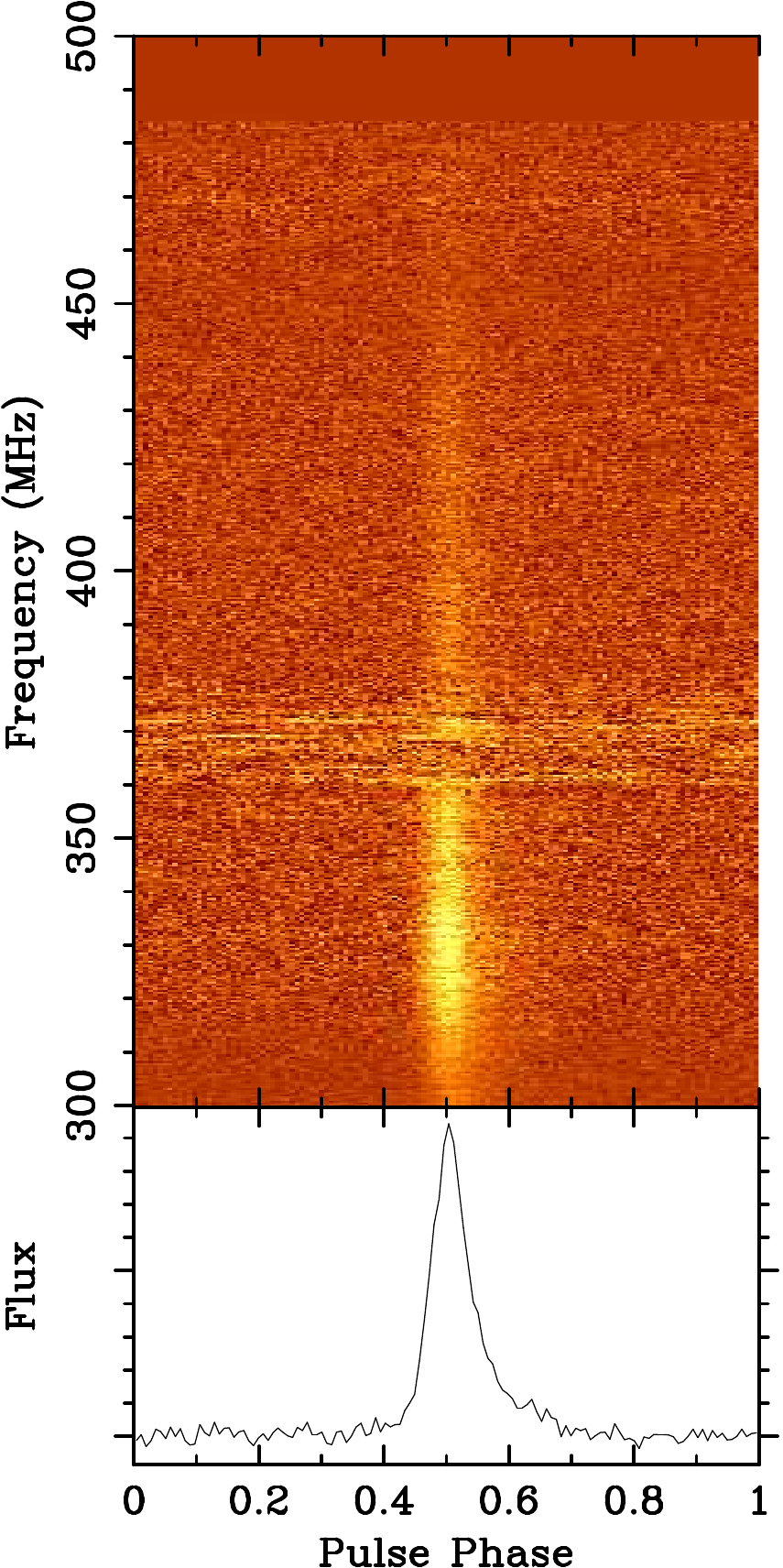}
\caption[]{Observation of NGC~1851A as taken with the 250$-$500~MHz (Band 3) receiver of the uGMRT on 2017 November 14 simultaneously in PA mode (left) and CDP mode (right).
Top panels:  intensity as a function of pulse phase ($x$-axis) and time ($y$-axis).
Bottom panels: corresponding integrated pulse profile.
Thanks to the coherent de-dispersion available in CDP mode, the pulse profile of NGC~1851A is much narrower than in PA mode, resulting in far more precise timing.
Radio frequency interference from the Mobile User Objective System (MUOS) satellite system is visible in the 360$-$380 MHz band.}
\label{fig:PA_CDP_comparison}
\end{figure}

\subsection{Observations}

We used the uGMRT to observe NGC~1851A at 20 different epochs, from 2017 April to 2018 March (Table
\ref{tab:NGC1851A_uGMRT_observations}). The observing strategy was designed with two main objectives in mind: a) improve the
measurements of the proper motion and spin-down parameters, and b) possibly measure additional PK parameters, such as the
Einstein delay and the Shapiro delay. For the first purpose, we carried out $\sim$40-min-long observations with a roughly
monthly cadence. For the second purpose, we carried out a single dense orbital campaign in 2017 May, during which the pulsar
was observed on ten different occasions within the $\sim18.8$ d of its orbit, with longer scans as the pulsar was
approaching periastron. All observations were made with the 14 antennas of the uGMRT central array, using the 250$-$500 MHz
receivers (uGMRT Band 3). After summing the two polarizations, the data were 16-bit digitized and recorded in search mode by
the GMRT Wideband Backend \citep[GWB,][]{Reddy+2017}, with a total bandwidth of 200 MHz centered at a frequency of 400 MHz. 

Until 2017 August, the data were taken in Phased Array (PA) mode only, with a sampling time of 81.92 $\mu$s and 2048,
$\sim$0.0977-MHz wide frequency channels. Given its dispersion measure (DM) of $\sim$52.14 \dmunit, the signal of NGC~1851A
had a dispersive smearing across each channel of 338 $\mu$s at the top (500 MHz) of the band, and of 1.565 ms at the bottom
(300 MHz) of the band. These translated into an effective resolution of 348 $\mu$s and 1.567 ms at the top and the bottom of
the band, respectively.

From 2017 September, the new real-time coherent de-dispersion (CDP) mode of the GWB became available. The last six
observations made from 2017 October to 2018 March were therefore made using the PA and CDP modes simultaneously. In CDP
mode, the observing band was divided into 512 frequency channels, which were coherently de-dispersed at the nominal DM of
NGC~1851A. Thanks to this, the chosen sampling time of 10.24 $\mu$s also corresponds to the effective time resolution of the
CDP data. The much higher quality provided by the CDP data over the PA data is evident from Fig.
\ref{fig:PA_CDP_comparison}, where we show a single observation of NGC~1851A as resulting from the two different modes. The
lack of intra-channel dispersive smearing in CDP mode results in a much narrower pulse profile. Its shape is thus much
closer to the intrinsic one, likely only slightly smeared by scattering.

\begin{table*}
\caption[List of the recent observations of NGC~1851A made with the uGMRT.]{List of the recent observations of NGC~1851A made with the uGMRT. All observations were carried out with the 250$-$500~MHz receiver (Band 3), GWB as backend, with 200 MHz of bandwidth. The dates and epochs reported are referred to the start time of the observation in UTC time standard. PA: Phased Array mode; CDP: Coherent De-dispersion mode.}
\label{tab:NGC1851A_uGMRT_observations}
\footnotesize
\centering
\renewcommand{\arraystretch}{1.0}
\vskip 0.1cm
\begin{tabular}{lclrllrc}
\hline
Date               & Epoch       & Mode          & Length   &    Sampling time    &  Number of      & Mean anomaly range  & Notes         \\
                   & (MJD)       &               & (min)    &    ($\mu$s)         &   channels      & (deg)  &      \\
\hline
2017 Apr 29      &    57872    &    PA         &    25    &    81.92            &    2048           &     74.00 $-$ \phantom{0}74.33   &    \\
2017 May 07        &    57880    &    PA         &    40    &    81.92            &    2048         &    223.75 $-$ 224.28    &    \\
2017 May 09        &    57882    &    PA         &    33    &    81.92            &    2048         &    265.41 $-$ 265.84     &    \\
2017 May 10        &    57883    &    PA         &    40    &    81.92            &    2048         &    284.54 $-$ 285.07 &  \\
2017 May 11        &    57884    &    PA         &    40    &    81.92            &    2048         &    303.76 $-$ 304.29    &    \\
2017 May 13        &    57886    &    PA         &   105    &    81.92            &    2048         &    340.82 $-$ 342.50    &    \\
2017 May 14        &    57887    &    PA         &   275    &    81.92            &    2048         &    357.45 $-$ 361.11    & Periastron passage   \\
2017 May 15        &    57888    &    PA         &    96    &    81.92            &    2048         &     17.83 $-$ \phantom{0}19.11   &    \\
2017 May 17        &    57890    &    PA         &    53    &    81.92            &    2048         &     58.58 $-$ \phantom{0}59.29     &    \\
2017 May 19        &    57892    &    PA         &    40    &    81.92            &    2048         &     96.79 $-$ \phantom{0}97.32     &    \\
2017 May 22        &    57895    &    PA         &    50    &    81.92            &    2048         &    150.22 $-$ 150.88     &    \\
2017 Jun 10       &    57914    &    PA         &    40    &    81.92            &    2048          &    155.93 $-$ 156.46    &    \\
2017 Jul 12       &    57946    &    PA         &    40    &    81.92            &    2048           &    45.63 $-$ \phantom{0}46.17    &    \\
2017 Aug 19     &    57984    &    PA         &    40    &    81.92            &    2048            &     53.23 $-$ \phantom{0}53.77  &    \\
2017 Oct 26    &    58052    &    PA / CDP   &    40    &    81.92 / 10.24    &    2048 / 512        &   273.68 $-$ 274.22       &    \\
2017 Nov 14   &    58071    &    PA / CDP   &    26    &    81.92 / 10.24    &    2048 / 512         &   294.65 $-$ 295.00      &    \\
2017 Dec 14   &    58101    &    PA / CDP   &    40    &    81.92 / 10.24    &    2048 / 512        &    148.65 $-$ 149.19      &    \\
2018 Jan 13    &    58131    &    PA / CDP   &    40    &    81.92 / 10.24    &    2048 / 512        &   359.94 $-$ 360.47      & Periastron passage   \\
2018 Feb 15    &    58164    &    PA / CDP   &    40    &    81.92 / 10.24    &    2048 / 512        &   271.41 $-$ 271.94      &    \\
2018 Mar 15    &    58192    &    PA / CDP   &    40    &    81.92 / 10.24    &    2048 / 512        &    88.18 $-$ \phantom{0}88.71      &    \\
\hline
\end{tabular}
\end{table*}

\subsection{Data reduction}

The newly taken uGMRT search-mode data were first folded with the \texttt{prepfold} routine of a slightly modified version
of the classic branch\footnote{The classic branch of \texttt{PRESTO} is currently the only version capable of dealing with
the GMRT data format.} of the \texttt{PRESTO}\footnote{\url{http://www.cv.nrao.edu/~sransom/presto}} \citep{Ransom2001}
pulsar search package, using the best NGC~1851A ephemeris available. The so produced \texttt{PRESTO} folded archives were
then converted into PSRFITS format using the \texttt{psrconv} routine of the
\PSRCHIVE\footnote{\url{http://psrchive.sourceforge.net}} pulsar software package \citep{van_Straten+2012} and then
carefully cleaned from radio frequency interference (RFI). All the PA and CDP archives were then separately summed together
to produce a high signal-to-noise (S/N) PA integrated profile, and a high-S/N CDP integrated profile, respectively. Both
profiles were then smoothed with a Wavelet transform (using the  \texttt{psrsmooth} routine of \PSRCHIVE) so as to obtain
two noise-free template profiles to be used with the PA and CDP datasets, respectively. The use of two different templates
for the different datasets is justified by the large differences in the observed profile shape of NGC~1851A in the PA and
CDP data (Fig. \ref{fig:PA_CDP_comparison}). Each noise-free template was then cross-correlated in the Fourier domain
\citep{Taylor1992} against the folded archives of the relative dataset to derive topocentric pulse times of arrival (ToAs).
Because the time stamps of the CDP data are known to have a positive offset of 0.67108864 seconds with respect to those of
the PA data, we took this difference into account  by subtracting the value from all the CDP ToAs\footnote{This was done by
using the \texttt{TEMPO}'s TIME statement in the ToA file.}. Also, the PA and CDP template profiles were aligned in phase,
using the profile peak as the reference point, so as to avoid the introduction of any additional phase offsets. In this way,
the PA and CDP ToAs were directly comparable and no arbitrary offset between the two datasets was needed\footnote{Although
the different profile shapes could also introduce an additional offset between the PA and CDP ToAs, the timing solutions
obtained with and without accounting for such an offset proved to be compatible within 1-$\sigma$.}.

The new uGMRT ToAs were then used to extend the pulsar ephemeris published by \cite{Freire+2007} to the present time, using
the {\tt TEMPO}\footnote{\url{http://tempo.sourceforge.net}} pulsar timing package. When doing so, the ToAs are first
referred to the Terrestrial Time standard of the Bureau International des Poids et Mesures (BIPM).  In order to subtract the
motion of the radio telescope around the Earth's centre, {\tt TEMPO} uses the International Earth Rotation Service tables
and the known coordinates of the telescope. The Earth's motion relative to the Solar System barycentre (SSB) was also
subtracted by {\tt TEMPO} using the DE 430 Solar System ephemeris derived by the Jet Propulsion Laboratory 
\citep{Folkner+2014}. The resulting timing parameters are presented in Barycentric Dynamical Time (TDB).

We used two of the {\tt TEMPO} orbital models to analyze the data, all based on the description of
\cite{1985AIHS...43..107D,1986AIHS...44..263D}. The first is the ``DDFWHE'' model \citep{2016ApJ...829...55W}, which is
based on theory-independent ``DD'' model, but with the orthometric parameterization of the Shapiro delay described by
\cite{2010MNRAS.409..199F}. The second is the ``DDK'' model, which will later be used in the Bayesian analysis outlined in
section \ref{sec:masses}. This is, again, based on the DD model but takes into account the kinematic effects described by
\cite{1995ApJ...439L...5K,1996ApJ...467L..93K} and was implemented in {\tt TEMPO} by \cite{2003ASPC..302...65V}.

\section{Results}
\label{sec:results}

\begin{figure*}
\begin{center}
\includegraphics[width=0.9\textwidth, angle=0]{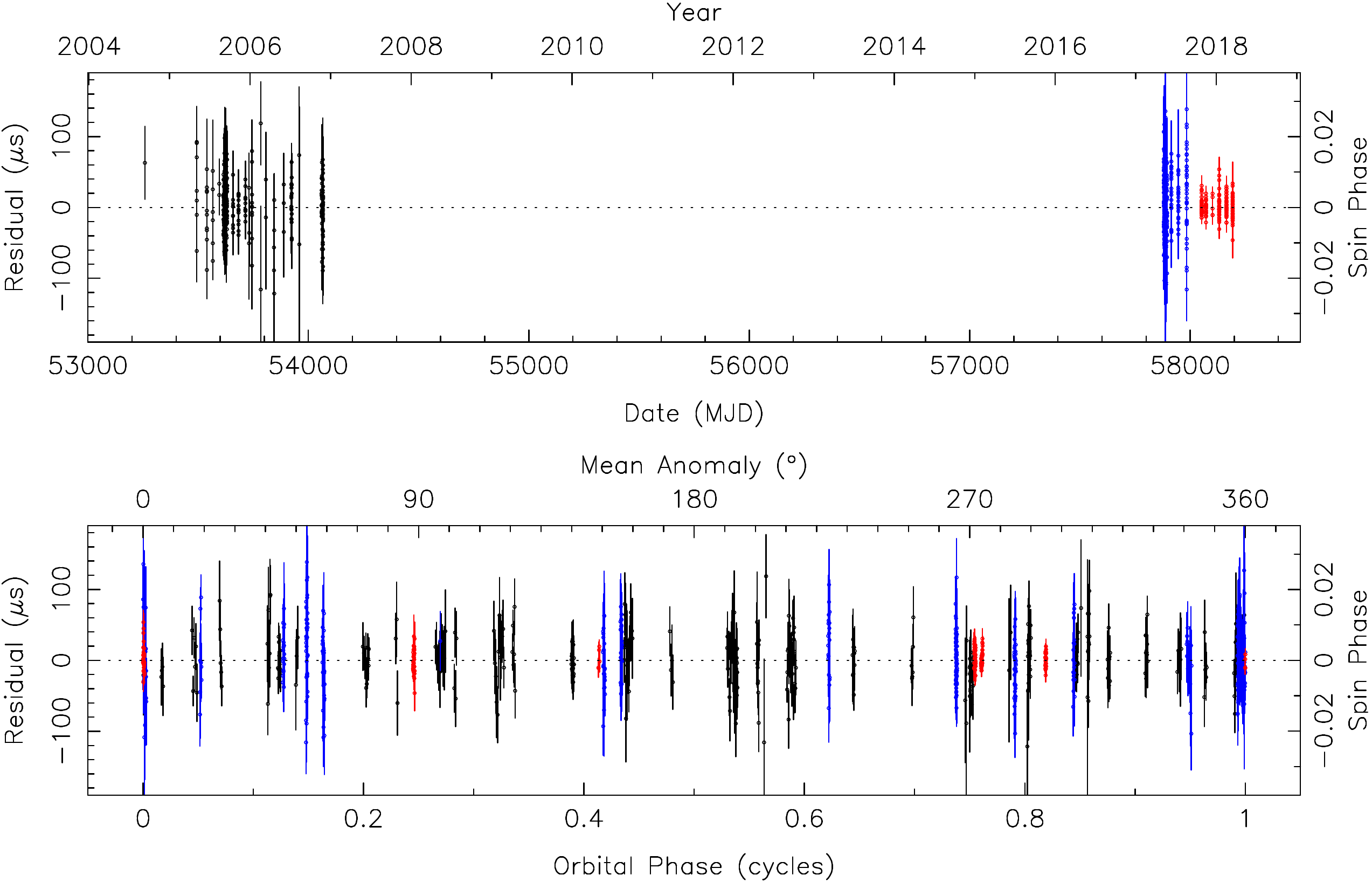}
\end{center}
\caption{Timing residuals for NGC~1851A, obtained with the DDFWHE timing
solution listed in Table~\ref{table:solution}. {\em Top}: ToA residuals as a function of the epoch, with the 10-year gap in observations evident.
{\em Bottom:} ToA residuals as a function of the
orbital phase, with phase 0 denoting periastron.
The residual 1-$\sigma$ uncertainties are indicated by
vertical error bars. Black indicates the earlier GBT timing, blue the 
uGMRT data taken in phased array (PA) mode, and red the uGMRT data in
coherent de-dispersion (CDP) mode. Note the marked improvement of the
CDP data compared to the PA mode.}
\label{fig:residuals}
\end{figure*}

\begin{table*}
\begin{center}{
\caption{Timing parameters for PSR~J0514$-$4002A\label{table:solution}}
\begin{tabular}{l c}
\hline
  \hline
  \multicolumn{2}{l}{Observation and data reduction parameters}\\
  \hline
  Reference Epoch (MJD) \dotfill & 53623.1551 \\
  Span of timing data (MJD) \dotfill & 53258 $-$ 58192 \\
  Number of ToAs \dotfill & 939 \\
  Solar wind parameter, $n_0$ (cm$^{-3}$) \dotfill & 10 \\
  Overall residual rms ($\mu s$) \dotfill & 20.9 \\
  RMS residual for GBT data ($\mu s$) \dotfill & 30.2 \\
  RMS residual for GMRT PA data ($\mu s$) \dotfill & 36.1 \\
  RMS residual for GMRT CDP data ($\mu s$) \dotfill & 11.4 \\
  $\chi^2$ \dotfill & 927.31 \\
  Reduced $\chi^2$ \dotfill & 1.008  \\
  \hline
  \multicolumn{2}{l}{Astrometric and spin parameters}\\
  \hline
  Right ascension, $\alpha$ (J2000) \dotfill & 05:14:06.69271(20) \\    
  Declination, $\delta$ (J2000) \dotfill & $-$40:02:48.8930(19) \\      
  Proper motion in $\alpha$, $\mu_\alpha$ ($\mathrm{mas\;yr^{-1}}$) \dotfill & 5.19(22) \\     
  Proper motion in $\delta$, $\mu_\delta$ ($\mathrm{mas\;yr^{-1}}$) \dotfill & $-$0.56(25) \\  
  Parallax, $\varpi$ (mas) \dotfill & 0.0826 \\
  Spin frequency, $\nu$ (Hz) \dotfill & 200.37770740535(10) \\
  First derivative of $\nu$, $\dot{\nu}$ ($10^{-17}\, \rm Hz \, s^{-1}$) \dotfill & $-$2.8(5) \\
  Second derivative of $\nu$, $\ddot{\nu}$ ($10^{-24}\, \rm Hz \, s^{-2}$) \dotfill & $-$1.533(27) \\
  Dispersion measure, DM ($\rm pc\,  cm^{-3}$) \dotfill & 52.14016(37) \\
 \hline
  \multicolumn{2}{l}{Binary parameters}\\
  \hline
  Orbital period, $P_{\rm b}$ (days) \dotfill & 18.785179217(19) \\
  Projected semi-major axis, $x$ (lt-s) \dotfill & 36.29028(27) \\
  Epoch of periastron, $T_0$ (MJD) \dotfill & 53623.15508797(35)  \\
  Orbital eccentricity, $e$ \dotfill & 0.8879771(11) \\
  Longitude of periastron, $\omega$ ($^\circ$) \dotfill & 82.3402(31) \\
  Rate of advance of periastron, $\dot{\omega}$ ($\deg\, \rm yr^{-1}$) \dotfill & 0.0129592(16)  \\
  Einstein delay, $\gamma$ (s) \dotfill & 0.0216(9) \\
  Derivative of $P_{\rm b}$, $\dot{P}_{\rm b}$ ($10^{-12}$ s s$^{-1}$) \dotfill & 22(9) \\
  Orthometric amplitude of Shapiro delay, $h_3$ ($\mu$s) \dotfill & 0.2(13) \\
  Orthometric ratio of Shapiro delay, $\varsigma$ \dotfill & $0.498^{\rm a}$ \\
  \hline
  \multicolumn{2}{l}{Derived parameters}\\
  \hline
  Magnitude of proper motion, $\mu$ ($\rm mas \, yr^{-1}$) \dotfill & 5.22(22) \\
  PA of proper motion, $\Theta_{\mu}$ ($\deg$, J2000) \dotfill & 96.2(28) \\
  PA of proper motion, $\Theta_{\mu}$ ($\deg$, Galactic) \dotfill & 14.8(28) \\  
  Spin period, $P$ (ms) \dotfill & 4.9905751141121(24) \\
  Spin period derivative, $\dot{P}$ ($10^{-22}$~s s$^{-1}$) \dotfill  & 7.0(13) \\
  Mass function, $f$ ($\rm M_{\odot}$) \dotfill & 0.1454196(33) \\
  Orbital inclination ($\deg$) \dotfill & $52$ \\
  Total mass, $M_{\rm tot}$ ($\rm M_{\odot}$) \dotfill & $2.4730(6)^{\rm b}$ \\
  Pulsar mass, $M_{p}$ ($\rm M_{\odot}$) \dotfill & $1.25^{+0.06}_{-0.05}$ \\
  Companion mass, $M_c$ ($\rm M_{\odot}$) \dotfill & $1.22^{+0.05}_{-0.06}$\\
  Angular distance from cluster center, $\theta_\perp$ (arcmin)   & 0.0784\\
\hline
\multicolumn{2}{l}{Notes. Timing parameters and 1-$\sigma$ uncertainties derived using {\sc tempo} in}\\
\multicolumn{2}{l}{Barycentric Dynamical Time (TDB), using the DE 430 Solar System ephemeris,}\\
\multicolumn{2}{l}{the Terrestrial Time (BIPM) timescale and the DDFWHE orbital model.}\\
\multicolumn{2}{l}{$^\textrm{a}$ Derived from $\dot{\omega}$ and $\gamma$ and held fixed (see section \ref{sec:shapiro}). $^\textrm{b}$ Derived from $\dot{\omega}$. }\\
\multicolumn{2}{l}{$d$ is the estimated distance to NGC~1851, its inverse is used for the parallax.}\\
\multicolumn{2}{l}{Estimate of $v_{\rm T}$, $\dot{P}_{\rm int}$ and derived parameters assume $d$.}
\end{tabular}
\vspace{-0.5cm}
}
\end{center}
\end{table*}

The timing parameters for NGC~1851A are presented in Table~\ref{table:solution}. The ToA residuals (calculated as observed
ToA $-$ prediction of the timing solution for the same rotation number) are displayed graphically in Fig.
\ref{fig:residuals}. These show no clear systematic trends, which suggests that the ephemeris in Table~\ref{table:solution}
accounts well for the spin and motion of the pulsar.

We will now discuss the astrometric, spin and binary parameters in this solution. However, before we proceed, we must remark
that some parameters, like the second spin frequency derivative, the proper motion and the orbital period derivative, are
still subject to change, showing significant differences with every new observation added. Therefore, their values must be
interpreted with caution; they will be discussed in more detail in a future publication after further timing provides stable
measurements for those parameters. Other parameters, particularly those used to derive the masses of the components, appear
to be much more robust. For this reason, the bulk of the discussion will be centered on the mass measurements.

The main new observational result in this section is the detection of the Einstein delay $\gamma$ (see
section~\ref{sec:gamma}). Another highlight is the detailed interpretation of $\gamma$, in particular the study of its
correlation with the rate of change of the projected semi-major axis of the pulsar orbit, $\dot{x}$. in
sections~\ref{sec:covariance} to \ref{sec:pm_gamma}.

\subsection{Proper motion: transverse velocity relative to NGC 1851} \label{sec:proper_motion}

\begin{figure}
  \includegraphics[width=\columnwidth]{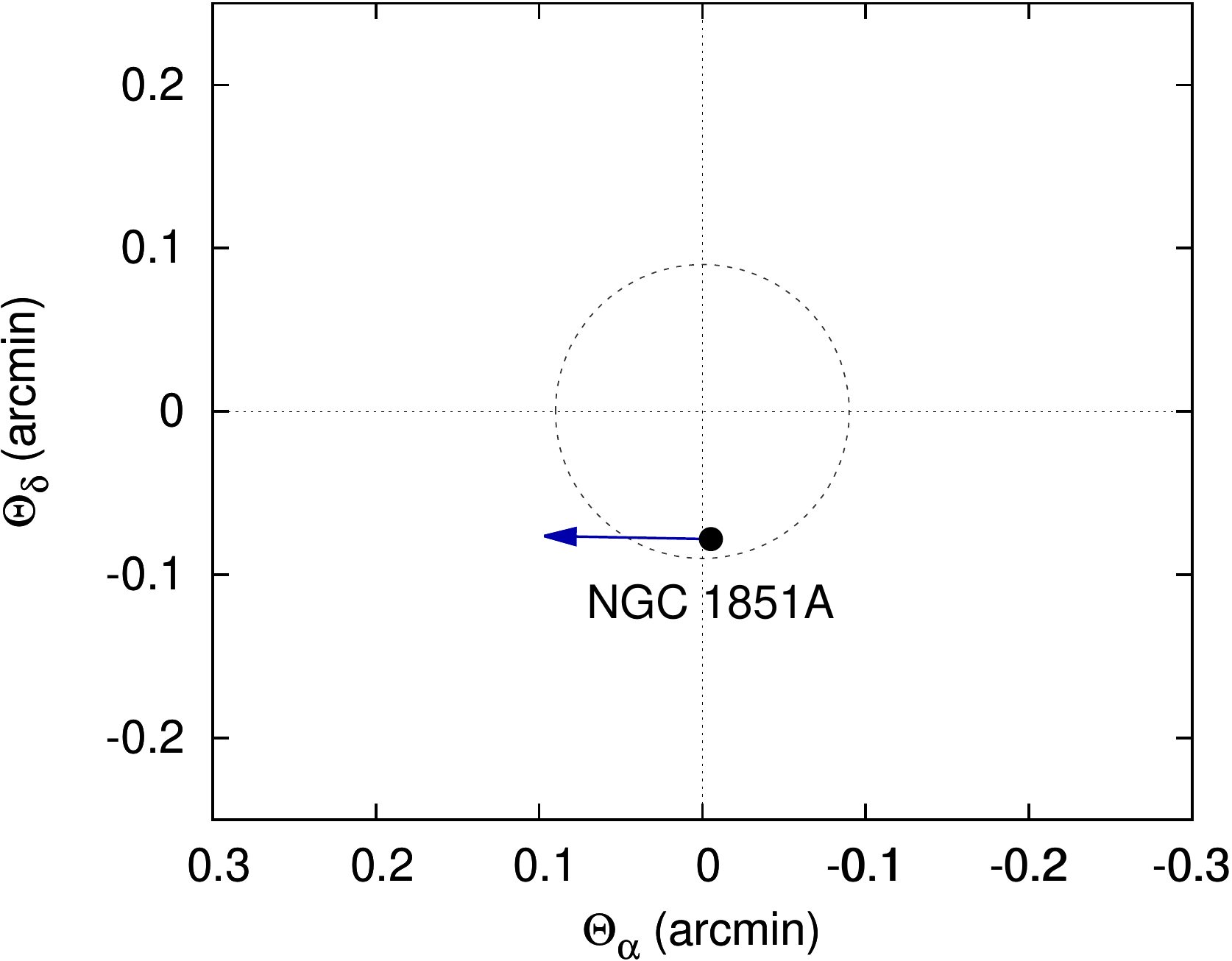}
  \caption[]{Position of NGC 1851A with respect to the nominal center of the cluster, located at $\alpha = 05^{\rm h}14^{\rm m}06.72^{\rm s}$ and $\delta=-40^\circ02\arcmin44.2\arcsec$ \citep{Gaia+2018}. The dashed circle shows the core radius of NGC 1851, which is $0.09 \arcmin$, according to \citet[][2010 Edition]{Harris1996}. The blue arrow indicates the direction of the projected motion of the pulsar relative to the cluster.}
  \label{fig:relative_propermotion}
\end{figure}

Contrary to the first timing solution published by \cite{Freire+2007}, which was limited by a $\sim$2-yr dataset, the much
longer time baseline spanned by our ToAs allowed us to precisely measure the pulsar's proper motion. The latter amounts to
$\mu_{\alpha} = +5.19 \pm 0.22~\pmunit$ in right ascension and $\mu_{\delta} = -0.56 \pm 0.25~\pmunit$ in declination.  
This can be compared to the astrometric measurement of the proper motion of the cluster as a whole, as recently published by
the GAIA Collaboration with their Data Release 2 \citep{Gaia+2018}. For NGC~1851 they report $\mu_{\alpha} = +2.1308 \pm
0.0037~\pmunit$ and $\mu_{\delta} = -0.6220 \pm 0.0040~\pmunit$. The motion of the pulsar relative to the cluster is
therefore $\Delta \mu_{\alpha} = +3.06\,\pm\,0.22~\pmunit$ and $\Delta \mu_{\delta} = +0.06  \pm 0.25~\pmunit$ and it is
graphically shown in Fig. \ref{fig:relative_propermotion}. Given the distance to NGC 1851 of $d = 12.1 \pm 0.2$~kpc
\citep{Gaia+2018}, this translates into a relative linear velocity of $\sim175 \pm 13$~km/s. This is more than four times
larger than the cluster's central escape velocity ($\sim42.9$~km/s, \citealt{Baumgardt_Hilker2018}), hence it would imply
that the pulsar is not bound to NGC~1851. However, we point out that this result is a direct consequence of the large
discrepancy between the proper motion of the pulsar, measured by radio timing, and that of the cluster, measured by the much
more precise GAIA's astrometry. Such a discrepancy may be due to covariances between the proper motion and the pulsar's
spin-down parameters, as the latter can be heavily affected by the cluster's gravitational potential (see discussion in
Section \ref{sec:pdot}).  Considering the 10-yr gap in our timing data, and the fact that the recent one year of data was
taken with two different back-ends, it is too early to draw any firm conclusions. Further radio observations of NGC 1851A
over the next few years will be necessary before we are able to accurately measure higher order spin period derivatives,
which will in turn improve the measurement of the pulsar's proper motion.

\subsection{Keplerian orbital parameters}
\label{sec:massf}

As already discussed by \cite{Freire+2004} and \cite{Freire+2007}, NGC~1851A is, among binary MSPs, an unusually eccentric
system, $e \, = \, 0.8879771(11)$, the second highest after PSR J1835$-$3259A in NGC 6652, with $e \, = \, 0.968(5)$
\citep{DeCesar+2015}. Furthermore, it must have an unusually massive companion, since it has a high mass function, $f$:
\begin{equation} \label{eq:f} f \, = \, \frac{(M_c \sin i)^3}{M_{\rm tot}^2} \, = \, \frac{4 \pi^2}{T_{\odot}}
\frac{x^3}{P_{\rm b}^2} \, = \, 0.1454196(33) \, M_{\odot}, \end{equation} where $M_{\rm tot}$ is total system mass, $x$ is
the projection of the semi-major axis of the pulsar's orbit along the line of sight in light-seconds (lt-s), $P_{\rm b}$ is
the orbital period of the binary, and $T_{\odot} = G M_{\odot} c^{-3} = 4.925490947 \mu \rm s$ is a solar mass ($M_{\odot}$)
in time units, where $c$ is the speed of light and $G$ is Newton's gravitational constant.

The total mass $M_{\rm tot}$ was already presented by \cite{Freire+2007}, but in this
work we present a much more precise value, $M_{\rm tot}\, = \, \, 2.4730(6) \, \rm M_{\odot}$
(see section~\ref{sec:omdot}). Thus, from eq.~\ref{eq:f} we derive:

\begin{equation}
\label{eq:sini}
M_c \, = \, \frac{1}{\sin i}(f M_{\rm tot}^2)^{\frac{1}{3}} \, = \, \frac{0.96166(16)\, M_\odot}{ \sin i},
\end{equation}

this means that for the largest possible $\sin i$, $M_c$ has a minimum value of $0.96166(16) \, M_{\odot}$. The large mass
for the companion and large eccentricity indicate that the system is a product of a secondary exchange encounter, as already
pointed out by \cite{Freire+2007}.

Knowing $M_{\rm tot}$ fixes the sum of the semi-major axes of both components of the binary, also known as the orbital
separation, $a$, independently of the orbital inclination of the system. This results from Kepler's third law:

\begin{equation}
\label{eq:a}
a \, = \, c \left[ M_{\rm tot} T_{\odot} \left(\frac{P_{\rm b}}{2 \pi}  \right)^
2  \right]^{1/3}
\, = \, 2.79776(23) \, \times \, 10^{10}~\rm m,
\end{equation}
or 93.323(8) lt-s. We will need this value in some of the calculations below.
This also implies that there is a minimum value of $\sin i$, which we can
obtain from eq.~\ref{eq:sini} by assuming $M_p \, = \, 0$, $M_c \, = \, M_{\rm tot}$;
this is $\sin i \, > \, 0.3888$.

\subsection{Variation of the spin period}
\label{sec:pdot}

The observed variation of the spin period is given by:
\begin{equation}
\label{eq:pdot}
\left( \frac{\dot{P}}{P} \right)^{\rm obs} \, = \, \left( \frac{\dot{P}}{P} \right)^{\rm int} 
- \frac{\dot{D}}{D},
\end{equation}
where $\dot{P}_{\rm int}$ is the intrinsic spin-down of the pulsar and
$\dot{D}$ is the variation of the Doppler shift factor $D$.
As already noticed by \cite{Freire+2007}, the observed $\dot{P}$ is extremely small,
our updated value is $7.0 \, \pm \, 1.3\, \times \, 10^{-22} \, \rm s \, s^{-1}$, thus
$(\dot{P}/P)^{\rm obs} \, = \, 1.41(26) \, \times \, 10^{-19} \, \rm s^{-1}$.

The variation of the Doppler shift factor ($D$) consists of an acceleration term proportional to the distance, $d$ and the
square of the total proper motion, $\mu$ \citep{1970SvA....13..562S}, another term due to the effect of the difference in
the Galactic accelerations of the pulsar's system and the Solar System, projected along the direction from the pulsar to the
Earth, $a_l$ \citep{1991ApJ...366..501D} plus, in this case, the (unknown) acceleration of the system in the gravitational
field of the cluster, $a_g$:

\begin{equation}
\frac{\dot{D}}{D} \, \equiv - \frac{\mu^2 d + a_l + a_g}{c},
\end{equation}
where, again, $c$ is the speed of light. In order to separate it from the $\dot{P}_{\rm int}$,
\cite{Freire+2017} used a similar expression for the orbital period:
\begin{equation}
\label{eq:pbdot}
\left( \frac{\dot{P}_{\rm b}}{P_{\rm b}} \right)^{\rm obs} \, = \,
\left( \frac{\dot{P}_{\rm b}}{P_{\rm b}} \right)^{\rm int} - \, \frac{\dot{D}}{D}.
\end{equation}
Subtracting equation (\ref{eq:pbdot}) from equation~(\ref{eq:pdot},) we obtain 
a result that does not depend on the acceleration or the proper motion of the system:
\begin{equation}
\label{eq:pdot_pbdot}
\Delta \, \equiv \, \left( \frac{\dot{P}}{P} \right)^{\rm obs} \, - \,
\left( \frac{\dot{P}_{\rm b}}{P_{\rm b}} \right)^{\rm obs} \, = \,
\left( \frac{\dot{P}}{P} \right)^{\rm int} \, - \,
\left( \frac{\dot{P}_{\rm b}}{P_{\rm b}} \right)^{\rm int}.
\end{equation}

\cite{Freire+2017} then assumed that $\dot{P}_{\rm b}^{\rm int}$ is small to obtain estimates of the intrinsic $\dot{P}^{\rm
int}$ for several MSPs in the globular cluster 47~Tucanae; these showed that they are very similar to the MSPs in the
Galactic disk.

For NGC~1851A, we cannot make this assumption. The reason is that if we evaluate the left side of
equation~(\ref{eq:pdot_pbdot}), we obtain $\Delta \, = \, -1.4(5) \, \times \, 10^{-17} \rm\,  s^{-1}$. If  $\dot{P}_{\rm
b}^{\rm int}$ could be ignored, then $\Delta$ would be positive, since for a rotation-powered pulsar $\dot{P}^{\rm int}$ is
always positive. The fact that it is negative means that it is possibly being compensated by a larger, positive
$\dot{P}_{\rm b}^{\rm int}$. Because this analysis is independent of the system's acceleration, the negative $\Delta$
cannot be explained by the acceleration of the system in the gravitational field of a nearby star, or the cluster's. We
discuss this in the following section.

Therefore, in order to estimate $\dot{D}/D$, we assume two extreme characteristic ages for the pulsar, 0.5 and 10 Gyr; these
bracket the characteristic ages of most known MSPs. Using these ages, we get values for $\dot{P}_{\rm int}$ of $1.6 \,
\times \, 10^{-19} \, \rm s \, s^{-1}$ and $7.9 \, \times \, 10^{-21} \, \rm s \, s^{-1}$ respectively. From
equation~(\ref{eq:pdot}), we then obtain for $\dot{D}/D$ the  extreme values of $3.2 \, \times \, 10^{-17} \, \rm s^{-1}$
and  $1.44 \, \times \, 10^{-18} \, \rm s^{-1}$. After subtraction of the proper motion and Galactic acceleration terms, we
obtain a very small line-of-sight acceleration for this binary system. This does not introduce any useful constraints on
cluster mass models, for this reason we will not elaborate on it any further.

\subsection{Variation of the orbital period}
\label{sec:pbdot}

\begin{figure}
  \includegraphics[width=\columnwidth]{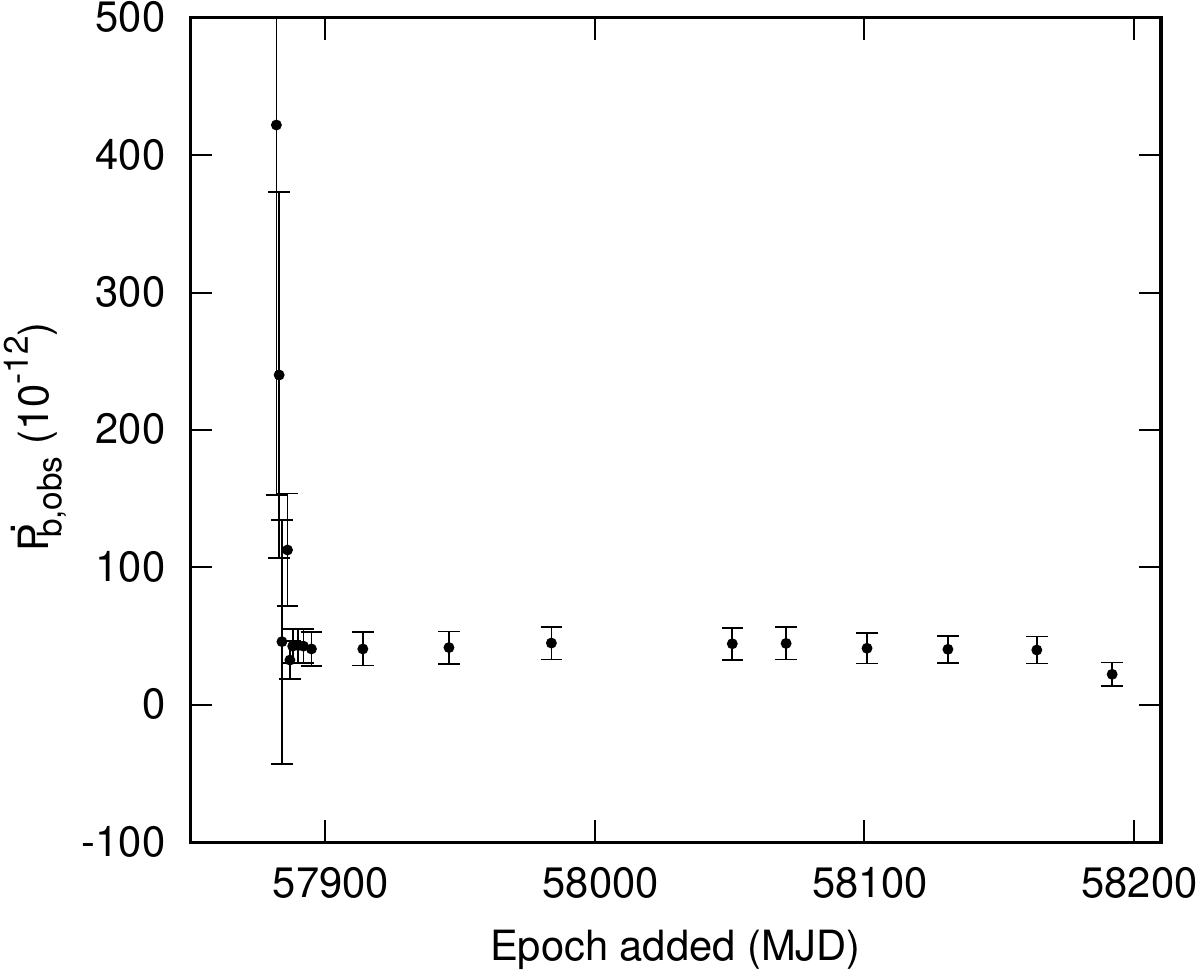}
  \caption[]{Orbital period derivative ($\dot{P}_{\rm b, obs}$) of NGC~1851A measured using the DDFWHE binary model as a function of epoch added from uGMRT dataset (the GBT dataset is also always included in the fit). The last point, which is the one derived using the whole GBT+uGMRT dataset, is the value reported in Table~\ref{table:solution}.} 
  \label{fig:pbdot_vs_epoch}
\end{figure}

According to \cite{2004hpa..book.....L}, the observed variation of the orbital period is given by: \begin{eqnarray}
\label{eq:pbdot_null} \left( \frac{\dot{P}_{\rm b}}{P_{\rm b}} \right)^{\rm int} & = & \left( \frac{\dot{P}_{\rm b}}{P_{\rm b}}
\right)^{\rm obs}\, + \, \frac{\dot{D}}{D} \\ & = & \left( \frac{\dot{P}_{\rm b}}{P_{\rm b}} \right)^{\rm GW} \, + \, \left(
\frac{\dot{P}_{\rm b}}{P_{\rm b}} \right)^{\rm \dot{m}}\, + \, \left( \frac{\dot{P}_{\rm b}}{P_{\rm b}} \right)^{\rm T} \nonumber.
\end{eqnarray} Depending on the assumption above for $\dot{P}^{\rm int}$ and $\dot{D}/D$ we get  for this sum a range of values from
1.6(5) to $4.5(5) \, \times \, 10^{-17} \rm \s^{-1}$. This implies that the result appears to be at least 3-$\sigma$ significant.
However, we have noticed already that the value of $\dot{P}_{\rm b,\, obs}$ has not fully stabilized yet. As shown in Fig.
\ref{fig:pbdot_vs_epoch}, as we add more and more epochs to our dataset, its positive value keeps changing and ultimately tends to
decrease. For this reason, we will only present a brief  discussion of this effect below. In particular, we look at the individual
terms and discuss whether they could yield a large, positive $\dot{P}_{\rm b, \, int}$ or not.

The first term on the second line of 
equation~(\ref{eq:pbdot_null}) is due to loss of orbital energy caused by the emission of 
quadrupolar gravitational waves. Assuming the masses obtained in section~\ref{sec:masses},
this is given in GR by \citep{1964PhRv..136.1224P}:
\begin{eqnarray}
\dot{P}_{\rm b, GR} & = & - \frac{192 \pi}{5} T_{\odot}^{5/3} \left(\frac{P_{\rm b}}{2 \pi} \right)^{-5/3} f(e) \frac{M_{\rm p} M_c}{M_{\rm tot}^{1/3}}\\
f(e) & = & \frac{1}{(1 - e^2)^{7/2}} \left( 1 + \frac{73}{24} e^2 + \frac{37}{96} e^4 \right).
\end{eqnarray}
For the masses of the system, as they are determined in section~\ref{sec:masses}, we obtain
$\dot{P}_{\rm b, GR} \, = \,  -0.155 \, \times \, 10^{-12} \, \rm s \, s^{-1}$, thus
$\dot{P}_{\rm b, GR} / P_{\rm b} \, = \,  -9.6 \, \times \, 10^{-20} \, \rm s \, s^{-1}$.
This is two orders of magnitude smaller than $\Delta$, thus it does not explain the anomalous $\Delta$ we
observe. 

The second term in the second line of equation~(\ref{eq:pbdot_null}) is caused by mass loss
from the system. Assuming, as \cite{1991ApJ...366..501D} did, that this is dominated
by the loss of rotational energy for the pulsar, this is given by :
\begin{eqnarray}
\label{eq:pdotmdot}
\left( \frac{\dot{P}_{\rm b}}{P_{\rm b}} \right)^{\rm \dot{m}} & = & 2 \frac{\dot{m}}{M_{\rm tot}} \, = \, 2 \frac{\dot{E}}{c^2 M_{\rm tot}}\\
 & = & \frac{8 \pi G}{T_{\odot} c^5} \frac{I}{M_{\rm tot}} \frac{\dot{P}_{\rm int}}{P^3} \, \simeq \, 2.3 \, \times\, 10^{-4} \left( \frac{\dot{P}}{P} \right)^{\rm int}
\end{eqnarray}
where $I \, \simeq \, 10^{38} \rm \, kg\, m^2$ is the moment of inertia of the pulsar.
We can see from the 
last identity that this term is extremely small compared to
$\dot{P}_{\rm int}/P$, which is very similar to $\dot{D}/D$ (equations~\ref{eq:pdot} and \ref{eq:pbdot}).
This means that this term cannot explain 
the observed $\Delta$ either.

However, if the companion is losing mass on its
own at a sufficiently large rate, that could cause the
observed $\Delta$. Indeed, if we assume that
$\Delta$ is caused by mass loss, we obtain:

\begin{align}
\label{eq:pbdot_mdot}
\left( \frac{\dot{P}_{\rm b}}{P_{\rm b}} \right)^{\dot{m}} & \simeq \left( \frac{\dot{P}_{\rm b}}{P_{\rm b}} \right)^{\rm obs} \, + \frac{\dot{D}}{D} \, > \, 1.55 \, \times \, 10^{-17}\, \rm s^{-1} \\
\dot{m} \,  & =  \, \frac{M_{\rm tot}}{2} \left( \frac{\dot{P}_{\rm b}}{P_{\rm b}} \right)^{\dot{m}}\\
\dot{m} & > 1.9 \times 10^{-17} M_{\odot} \, \textrm{s}^{-1}\, = 6.1 \times 10^{-10} M_{\odot} \,\textrm{yr}^{-1}
\end{align}

which is about $10^4$ times larger than the current mass loss rate for the Sun. Such mass loss rates do not generally occur for
compact objects (certainly not for NSs or heavy white dwarfs). However, as discussed by \cite{Freire+2007}, the lack of eclipses
rules out extended companions, such as main sequence star companions, and even more a giant companion. Therefore, a large mass loss
rate should not be expected.

Finally, the last term on the second line of equation~(\ref{eq:pbdot_null}) is caused by tidal dissipation. This might explain the
observed $\Delta$ if the companion were extended and rotated fast and in the same sense of the orbit, as in the Earth-Moon system.
However, since the companion does not appear to be extended, this is, again, an unlikely explanation.

Since the value of $\dot{P}_{\rm b}^{\rm obs}$ has not fully stabilized yet, there is a chance that none of these effects (mass loss
or tidal acceleration of the orbit) are real. To our knowledge, this effect has not been observed in any pulsars to date. Continued
timing with the coherent de-dispersion mode will quickly improve its precision and robustness and confirm the increase in the orbital
period or not.

\begin{figure*}
\centering
  \includegraphics[width=0.8\textwidth]{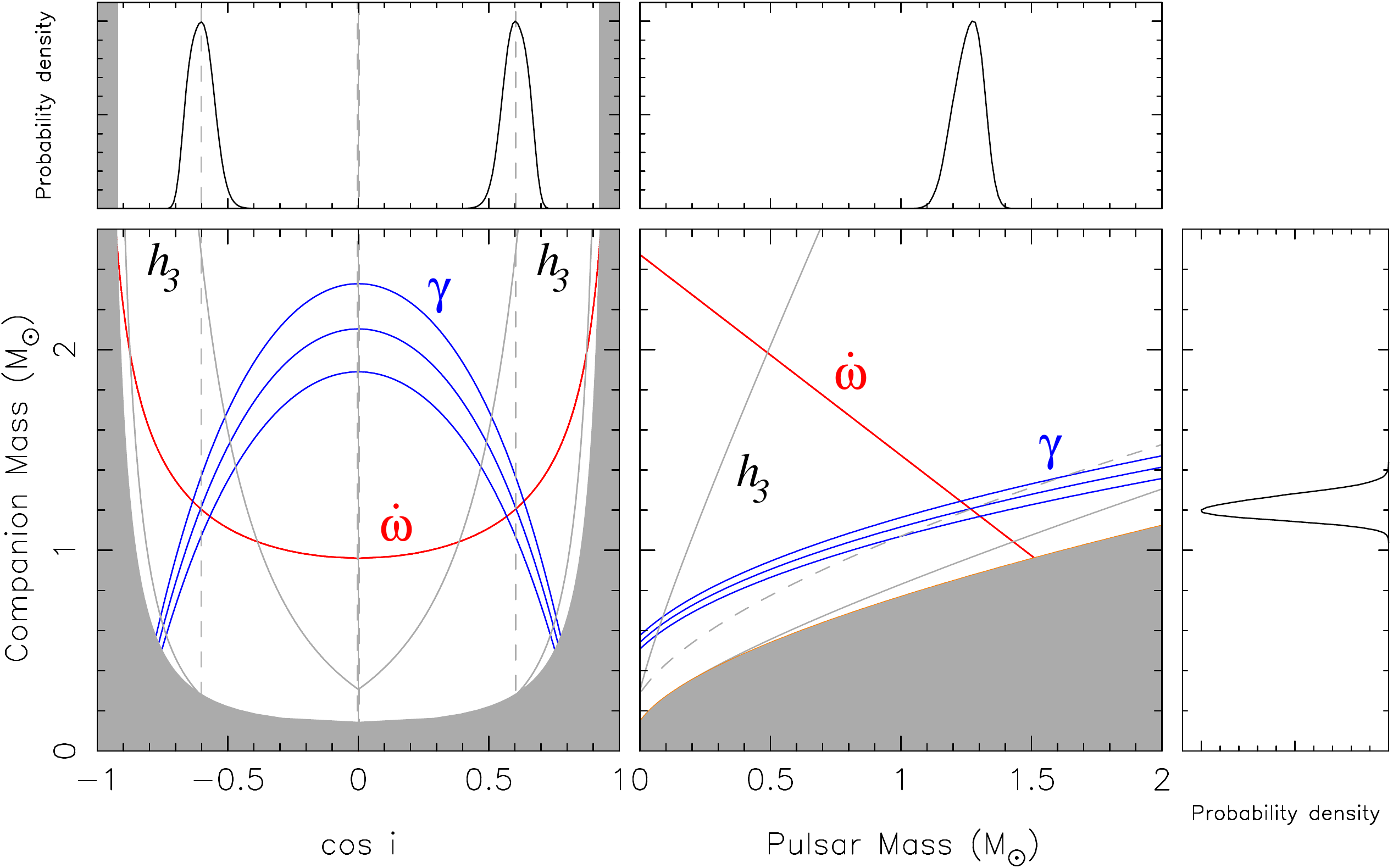}
  \caption[]{Mass constraints for PSR~J0514$-$4002A. In the main square plots, the
lines indicate the regions that are (according to general relativity)
consistent with the nominal and $\pm\, 1 \sigma$ measurements of $\dot{\omega}$
(solid red), $\gamma$ (solid blue) and $h_3$ (gray) obtained from
the DDFWHE model (see Table~\ref{table:solution}). For the $h_3$ estimate, we assumed
the value of $\varsigma$ marked by the dotted gray lines (section~\ref{sec:shapiro}); only the
nominal (near $M_{\rm p} \, =\, 0$) and $+ 1 \, \sigma$ lines are visible;
the latter excludes orbital inclinations near $90 \, \deg$. In the left plot, we display
$\cos i$ (for randomly inclined orbits this would have equal probability) versus the
companion mass ($M_c$); the gray region is excluded by the mass function of the
system - the pulsar mass ($M_{\rm p}$) must be larger than 0.
In the right plot, we display $M_{\rm p}$ versus $M_c$; the gray region is excluded
by the constraint $\sin i\, \leq\, 1$. The side panels display the 1-d pdfs
for $\cos i$ (top left), $M_{\rm p}$ (top right) and $M_c$ (right), normalized to the maximum (for details see section~\ref{sec:masses}). The distribution for $M_c$ is
derived from the distribution for $M_c$ using $M_{\rm p}\, = \, M_{\rm tot} \, - \, M_c$.}
  \label{fig:masses}
\end{figure*}

\subsection{Rate of advance of periastron}
\label{sec:omdot}

For NGC~1851A, the observed rate of advance of periastron, $\dot{\omega}_{\rm obs}$ is
measured very precisely:
$\dot{\omega}_{\rm obs} \, = \, 0.0129592(16)\, \deg \, \rm yr^{-1}$.
This is 25 times more precise and slightly larger than the value published by \cite{Freire+2007},
$\dot{\omega}_{\rm obs} \, = \, 0.01289(4)\, \deg \, \rm yr^{-1}$. As all 
measurements in this work, the latter's uncertainty
is a 68.3 \% confidence limit, equivalent to 1 $\sigma$ in a normal distribution.
Our new value is thus $1.7$-$\sigma$ larger than the earlier. This difference is
not statistically significant.

According to \cite{2004hpa..book.....L}, in the absence of
other massive objects near the binary, $\dot{\omega}_{\rm obs}$ is given by:
\begin{equation}
\dot{\omega}_{\rm obs} \, = \, \dot{\omega}_{\rm rel} + \dot{\omega}_{\rm k} + \dot{\omega}_{\rm SO} 
\end{equation}
The first term is caused by relativistic effects. Assuming general relativity (GR), we can estimate
the total mass of the binary, $M_{\rm tot}$ (in solar masses), from $\dot{\omega}_{\rm rel}$
and the Keplerian parameters $P_{\rm b}$ and $e$ \citep{rob38}
by inverting the well-known expression derived by \cite{1982ApJ...253..908T}:
\begin{equation}
\label{eqn:M}
M_{\rm tot}\, = \, \frac{1}{T_{\odot}} \left[ \frac{\dot{\omega}_{\rm Rel}}{3} (1- e^2) \right]^{\frac{3}{2}} \left(\frac{P_{\mathrm{b}}}{2\pi} \right)^{\frac{5}{2}},
\end{equation}
If $\dot{\omega}_{\rm rel}$ fully accounts for $\dot{\omega}_{\rm obs}$, then we can derive 
$M_{\rm tot} \, = \, 2.47298(45) \, \rm M_{\odot}$. This constraint is displayed by the
red lines in figure~\ref{fig:masses}.

However, $\dot{\omega}_{\rm Rel}$ does not fully account for the observations.
The second term in equation~(\ref{eq:omdot_k}), $\dot{\omega}_{\rm k}$, is given by \cite{1995ApJ...439L...5K},
here re-arranged as in \cite{2011MNRAS.412.2763F}:
\begin{equation}
\label{eq:omdot_k}
\dot{\omega}_{\rm k} \, = \, \frac{\mu}{\sin i} \cos \left( \Theta_{\mu} - \Omega \right),
\end{equation}
where $\Theta_{\mu}$ is the position angle of the proper motion
and $\Omega$ is the (unknown) position angle for the line of nodes.
Maximizing this contribution, i.e., setting $\cos \left( \Theta_{\mu} - \Omega \right)\, = \, \pm \, 1$
(and using the value for $\sin i$, from section~\ref{sec:gamma}),
we obtain $\dot{\omega}_{\rm k} \, = \, \pm \, 1.82 \, \times \, 10^{-6} \, \rm \deg \, yr^{-1}$,
which is very similar to the measurement uncertainty. Thus the assumption that
$\dot{\omega}_{\rm obs}$ is caused by relativistic effects is mostly warranted,
but $\dot{\omega}_{\rm K}$ is already having an influence on the uncertainty of the
measurement of the total mass of the binary.

The last term, $\dot{\omega}_{\rm SO}$, has not yet been detected in any
binary pulsar, so we will for now assume it does not contribute significantly.

As we will see in section~\ref{sec:masses}, we have at the moment no way of
measuring $\Omega$. Therefore, $\dot{\omega}_{\rm k}$ cannot be evaluated,
beyond the lower and upper limits we have estimated. This means that the
uncertainty of $M_{\rm tot}$ has to be increased to take into account
the unknown contribution of $\dot{\omega}_{\rm k}$. Adding the maximum
value of $\dot{\omega}_{\rm k}$ to the uncertainty of $\dot{\omega}_{\rm obs}$,
we obtain an estimate for the uncertainty of the latter: 
$2.4 \, \times \, 10^{-6} \, \rm \deg \, yr^{-1}$.
This translates into a $M_{\rm tot}$ uncertainty of $6 \, \times 10^{-4}\, M_{\odot}$,
which is the uncertainty quoted in Table~\ref{table:solution}.

\subsection{Einstein delay}
\label{sec:gamma}

The main new result in this paper is the measurement of the Einstein delay,
$\gamma$. This measures the apparent slowdown of the rotation of the pulsar near
periastron relative to apastron. Assuming that it is solely an effect of GR (an assumption we discuss in detail below), 50\% of the effect is caused by the varying special-relativistic time dilation (which is caused by the varying velocity of the pulsar in its orbit) and 50\% by the varying gravitational redshift. 

Until now, this has been measured only for eccentric systems
with orbital periods of 10 hr (for PSR B15134+12, see \citealt{2014ApJ...787...82F})
or shorter, all of these being double neutron star systems.
The orbital period of NGC~1851A is 45 times larger than that of PSR B1534+12.
This detection was helped by the magnitude of the effect,
$\gamma \, = \, 21.6(9) \, \rm ms$; by far the largest ever measured in 
any binary pulsar.

In GR, this effect is related to the component masses by the equation
\begin{eqnarray}
\label{eq:gamma}
\gamma\, & = & \, \gamma_{\rm K} \frac{M_c (M_c + M_{\rm tot})}{M_{\rm tot}^{4/3}} \\
\gamma_{\rm K} & = & T_{\odot}^{2/3} e \left( \frac{P_{\rm b}}{2 \pi} \right)^{1/3},
\end{eqnarray}
these constraints are depicted by the blue lines in Fig.~(\ref{fig:masses}).
The reason for the large $\gamma$ of PSR~J0514$-$4002A has to do
with the $\gamma_{\rm K}$ term being larger for this pulsar,
a consequence of the large values for $e$ and $P_{\rm b}$.

If we already know $M_{\rm tot}$, we can determine the masses from $\gamma$ using
\begin{eqnarray}
\label{eq:gamma_masses}
M_c & = & \frac{1}{2} \left( \sqrt{ M_{\rm tot}^2 +
4 M_{\rm tot}^{4/3} \frac{\gamma}{\gamma_{\rm K}}} - M_{\rm tot} \right) \\
M_{\rm p} & = & M_{\rm tot} - M_c ;
\end{eqnarray}
for the $\gamma$ and $M_{\rm tot}$ measured for PSR~J0514$-$4002A, the result is $M_c \, = \, 1.207^{+0.037}_{-0.038} \, M_{\odot}$ and $M_{\rm p} \, = \, 1.266^{+0.038}_{-0.037} \, M_{\odot}$, assuming GR.

Inverting equation~(\ref{eq:sini}) we obtain $\sin i \, = \, 0.797$, which implies
either $i \, = \, 53.8\, \deg$ or $i \, = \, 127.2 \, \deg$.
These values are represented by the intersection of the red and blue lines in Fig.~\ref{fig:masses}. 

\subsection{Covariance of the Einstein delay with $\dot{x}$ for wide orbits}
\label{sec:covariance}

We now examine the assumption that the observed $\gamma$ is solely an
effect of GR. We start by examining why measurements of $\gamma$ for
wide binary pulsars have not been made to date, despite the (often)
very large expected values of $\gamma_{\rm K}$ and in some cases
$M_c$ as well.

As pointed out by \cite{1976ApJ...205..580B}, and later more explicitly by 
\cite{1998MNRAS.298..997W}, we cannot measure $\gamma$ for a
single orbit (even if measured with extreme precision) because
the effect is re-absorbed into the Keplerian parameters
$x$ and $\omega$.
If $x$ and $\omega$ are the ``real'' projected semi-major axis
and longitude of periastron for a particular binary, the
measurable, ``post-absorption'' quantities $x'$ and $\omega '$
are given, to very good approximation, by \citep{1998MNRAS.298..997W}:
\begin{eqnarray}
x' & = & x + \frac{\gamma}{\sqrt{1 - e^2}} \cos \omega \\
\omega ' & = & \omega - \frac{\gamma}{x \sqrt{1 - e^2}} \sin \omega,
\end{eqnarray}
for the parameters of NGC~1851A, we get $x' \, = \, 36.29656 \,$lt-s
and $\omega' \, = \, 82.2665 \, \deg$,
a difference of $0.00627(26)\,$lt-s and $-0.0737(30) \, \deg$
relative to the $x$ and $\omega$ in Table~\ref{table:solution}.
The uncertainties of the differences are calculated from the uncertainty of $\gamma$.

In order to measure $\gamma$, we must in effect measure
$x'$ for sufficiently spaced values of $\omega$. Given the large
$\dot{\omega}$ for the most compact and eccentric
double neutron star systems, such a
measurement is generally achievable for timing baselines of
a few years.
In the case of NGC~1851A, the $\dot{\omega}$
is only $0.0129592(16)\, \deg \, \rm yr^{-1}$, which means that a
full precession cycle lasts 27,779$\, \pm \,$3 years. This is, of course,
much longer than the timing baseline for this system, implying
that we can only observe the system at closely spaced values of $\omega$.

In such cases, we can only measure the current derivative of $x'$. Differentiating
the last equations we obtain:
\begin{eqnarray}
\label{eq:gamma_xdot}
\dot{x}' & = & \dot{x} - \frac{\gamma \dot{\omega}}{\sqrt{1 - e^2}} \sin \omega\\ 
\dot{\omega}' & = & \dot{\omega} - \frac{\gamma \dot{\omega}}{x \sqrt{1 - e^2}} \cos \omega
+ \frac{\gamma \dot{x}}{x^2 \sqrt{1 - e^2}} \sin \omega,
\end{eqnarray}
where we assumed that $e$ and $\gamma$ are constant.

Assuming specifically the DD model, we find that in
the equations above $e$ should be replaced by $e_{\theta}$, which is given by:
\begin{equation}
e_{\theta} \, = \, e (1 + \delta_\theta),
\end{equation}
where $\delta_{\theta}$ is a PK parameter, the relativistic deformation parameter.
In GR, this is expected to be $3.78 \, \times \, 10^{-7}$ for NGC 1851A. This parameter is
not separately measurable for this pulsar. The difference between $e$ and $e_\theta$
is so small that it can be ignored in the discussion that follows.

The difference between $\dot{\omega}$ and $\dot{\omega}'$,
$2.24 \, \times \, 10^{-6}\, \rm \deg \, yr^{-1}$, is similar to the uncertainty
on $\dot{\omega}$, for that reason we will also
ignore it for the time being (it is taken into account anyway when we fit for $\gamma$).

Since we only really measure $\dot{x}'$, we cannot separate the
instrinsic variation of the projected
semi-major axis ($\dot{x}$) from $\gamma$, both quantities are completely covariant (eq.~\ref{eq:gamma_xdot}).
Indeed, if we fit for both quantities in {\tt TEMPO}, we cannot determine either with any useful precision.

Since we have only fitted for $\gamma$ in our timing solution (not for $\dot{x}$),
the $\dot{x}'$ should be given by the $\gamma$ term in equation~(\ref{eq:gamma_xdot}):
\begin{equation}
\label{eq:x-dot_observed}
\dot{x}' \, = \, - 3.34(14) \, \times \, 10^{-13} \, \textrm{lt-s}\, \textrm{s}^{-1},
\end{equation}
where the uncertainty is derived from the uncertainty of $\gamma$.

We can test these expressions very easily by fitting the DDFWHE solution
with $\dot{x}$ instead of $\gamma$.
Doing this we obtain a fit with basically the same $\chi^2$ (927.29)
and the following parameters:
\begin{eqnarray}
\label{eq:x_xdot}
x & = & 36.29658(6)\, \textrm{lt-s}, \\ 
\omega & = & 82.266526(31) \, \deg,  \\
\dot{x} & = & -3.34(14)\, \times \, 10^{-13} \textrm{lt-s} \, \textrm{s}^{-1},
\end{eqnarray}
which agree within 1-$\sigma$, and exactly with our expectation for the ``absorbed'' values $x'$, $\omega'$ and $\dot{x}'$ respectively.

If the intrinsic $\dot{x}$ is small compared to $\dot{x}'$ in
equation~(\ref{eq:gamma_xdot}) (or if it can be determined
independently with a precision that is small compared to $\dot{x}'$),
then we can measure $\gamma$ and use it to determine the masses.
If not, then it becomes impossible to measure $\gamma$ and determine reliable masses from it.
This is a general condition that must be evaluated before attempting to determine $\gamma$ for any wide
binary system. We estimate $\dot{x}$ in the next section.

However, before we proceed, we remark that equation~(\ref{eq:gamma_xdot}) implies that, for some wide binaries (those with $\omega$ close to $0 \deg$ or $180 \deg$)
the $\sin \omega \, = \, 0$ term makes it virtually impossible to
measure $\gamma$ for those binaries, at least while that $\omega$ configuration
persists (which can be many thousands of years).
One of the factors that allows the measurement of
$\gamma$ for NGC~1851A is the favourable $\omega$ of $82.34 \, \deg$, which
almost maximizes the possible contribution of $\gamma$ to $\dot{x}'$.

\subsection{Variation of the projected semi-major axis}
\label{sec:x-dot}

Since we have no independent way of measuring $\dot{x}$, it is very important to carefully estimate it. We do this in this section within the framework of GR.

According to \cite{2004hpa..book.....L}, the change in $\dot x$ can be written, in the absence of any massive objects in the vicinity of the binary, as:
\begin{equation}
\label{eq:xdot}
\left( \frac{\dot x}{x} \right)\,  = \,
\left( \frac{\dot x}{x} \right)^{\rm k} \, + \,
\left( \frac{\dot x}{x} \right)^{\rm GW} \, + \,
\frac{d \epsilon_A}{dt} \, - \,
\frac{\dot{D}}{D} \, + \,
\left( \frac{\dot x}{x} \right)^{\dot{m}} + \,
\left( \frac{\dot x}{x} \right)^{\rm SO}
\end{equation}
The first term is caused by the changing geometry due to
the motion of the system relative to the Earth and it is given by \citep{1995ApJ...439L...5K}:
\begin{equation}
\label{eq:xdotk}
\left( \frac{\dot {x}}{x} \right)^{\rm k} = \mu \cot i \sin ( \Theta_\mu - \Omega),
\end{equation}
where we have, again, re-written the terms as in \cite{2011MNRAS.412.2763F},
except for the latter's negative sign; the reason for this is that the system we use to measure $\Omega$ and $i$ should be a right-handed system. Using the most likely value of $\sin i$ from section~\ref{sec:gamma}, we obtain for
this term a maximum and minimum values (corresponding to $\sin(\Theta_\mu - \Omega) \,= \, \pm 1$) of
$\pm\, 6.05 \, \times \, 10^{-16} \, \rm s^{-1}$.
This implies $\dot{x}_{\rm k} \, = \, \pm \, 2.2 \, \times \, 10^{-14} \, \rm lt$-$\rm s\, s^{-1}$.
This is about 6.5\% of $\dot{x}'$. Therefore, any computation of the component
masses will have to take this effect into account.

The second term is from the decrease of the size of the orbit caused by
gravitational wave emission; this is given by
\begin{equation}
\label{eq:adot}
\left( \frac{\dot{x}}{x} \right)^{\rm GW} \, = \, \frac{2}{3} \frac{\dot{P}_{\rm b, GW}}{P_{\rm b}} \, = \, -6.4 \times \, 10^{-20} \, \rm  s^{-1},
\end{equation}
i.e., $\dot{x}_{\rm GW} \, = \, -2.3 \, \times \, 10^{-18}  \rm lt$-$\rm s\, s^{-1}$.
This is four orders of magnitude smaller than $\dot{x}_{\rm k}$.

The third term, caused by aberration, is proportional to the geodetic precession rate for
the pulsar. This is given by \cite{1975PhRvD..12..329B} as:
\begin{equation}
\Omega_{\rm geod} \, = \, \left( \frac{2 \pi}{P_{\rm b}} \right)^{5/3}
T_{\odot}^{2/3} \frac{1}{1 - e^2} \frac{M_c (4 M_{\rm tot} - M_c) }{2 M_{\rm tot}^{4/3}},
\end{equation}
assuming the mass values derived in section~\ref{sec:masses} and the Keplerian
parameters of the system, we obtain
$\Omega^{\rm geod} \, = \, 0.0037 \, \deg \, \rm yr^{-1}$.
The aberration term is proportional to the latter \citep{1992PhRvD..45.1840D}:
\begin{equation}
\frac{d \epsilon_A}{dt}\, =\, \frac{P}{P_b} \frac{\Omega^{\rm geod}}{\sqrt{1 - e^2}}
\frac{\cot \lambda \sin 2 \eta + \cot i \cos \eta}{\sin \lambda},
\end{equation}
where $\eta$ and $\lambda$ are the polar coordinates of the pulsar's spin.
For NGC~1851A
the non-geometric factors (the first two fractions in the equation above)
amount to $1.37 \, \times \, 10^{-20} \rm \, s^{-1}$, i.e.,
the variation of $x$ caused by this term is about
$5.0 \, \times \, 10^{-19}\, \rm lt$-$\rm s\, s^{-1}$. This is more than four orders of magnitude smaller than $\dot{x}_{\rm k}$.

The fourth term is caused by the variation of the Doppler shift. From the assumption in section~\ref{sec:pbdot} of a characteristic age larger than 0.5 Gyr, it was deduced that
$\dot{D} / D \, < \, 3.2\, \times\, 10^{-17}\, {\rm s}^{-1}$,
i.e., its contribution to $\dot{x}$ is
$-1.1 \, \times \, 10^{-15}\, \rm lt$-$\rm s\, {\rm s}^{-1}$. This is
one order of magnitude smaller than $\dot{x}_{\rm k}$.

The fifth term can be derived from
$\dot{P}_{\rm b}^{\dot{m}}$ being given by equation~(\ref{eq:pbdot_mdot}).
If we use in that equation the upper estimate of
$\dot{D}/D$, as discussed in section~\ref{sec:pdot}, then we get
$(\dot{P}_{\rm b} / P_{\rm b})^{\dot{m}} \, < \, 4.6 \, \times \, 10^{-17} \, \rm s \, s^{-1}$.
Using equation~(\ref{eq:adot}),  we obtain
$(\dot{x} / x)^{\dot{m}}\, = \, 3.04 \, \times \, 10^{-17} \rm \, s^{-1}$,
i.e., $\dot{x}_{\dot{m}} \, = \, 1.1 \, \times \, 10^{-15} \rm lt$-$\rm s\, s^{-1}$.
This is one order of magnitude smaller than $\dot{x}_{\rm k}$, furthermore,
it is of a sign opposite to that of the contribution from $\dot{D} / D$ and
of very similar magnitude.

The sixth and last term, $\dot{x}^{\rm SO}$, has two contributions:
the relativistic spin-orbit coupling, also known as the Lense-Thirring effect ($\dot{x}_{\rm LT}$),
caused by the rotation of the
pulsar or the companion, and the classical spin-orbit coupling ($\dot{x}_{\rm QM}$).
Generally, for main sequence stars $\dot{x}_{\rm QM}$ is much larger than
$\dot{x}_{\rm LT}$, for NSs the opposite is true, and
for white dwarfs (WDs) both terms are roughly similar. 

For the Lense-Thirring effect, we have \citep{1992PhRvD..45.1840D}:
\begin{equation}
 \label{eq:LT}
 \dot{x}_{\rm LT}\, \simeq\, - x\frac{GS_{\rm A}}{c^2 a^3 (1-e^2)^{3/2}}\left(2 
+ \frac{3 M_B}{2 M_A}\right) \cot \textit{i} \sin{\delta_{\rm A}} \sin{\Phi^0_{\rm A}} 
 \end{equation} 
 where $S_{\rm A} = I_A \Omega_A$ is the rotational angular momentum of component A,
 $I_A$ is that component's moment of inertia and $\Omega_A$ is that component's
 angular frequency, $\delta_{\rm A}$ and $\Phi^0_{\rm A}$ are angles that determine
 the alignment of the rotation of component A relative to the orbit and  $a$ is the orbital separation calculated in equation~(\ref{eq:a}).

We now evaluate this term for the pulsar. For NSs, the moment of inertia is generally assumed to be of the order of $10^{38} \, \rm kg \, m^2$. For the pulsar, the angular frequency is well known,
$\Omega_{\rm p}\, = \, 2 \pi / P = 1259.01 \, \rm rad\, \s^{-1}$, thus 
$S_{\rm p} \, \sim \, 1.25 \, \times \, 10^{41} \, \rm kg \, m^2 \, s^{-1}$.
Therefore,
\begin{equation}
\dot{x}_{\rm LT}\, \simeq\, - 3 \, \times\, 10^{-15}  \sin{\delta_{\rm P}} \sin{\Phi^0_{\rm P}} \rm \, lts\ s^{-1},
\end{equation}
which is one order of magnitude smaller than the kinematic term $\dot{x}_{\rm K}$.
This will therefore have no impact on the mass measurements.

If the companion is a NS, then the same calculation can be made, except for the
lack of knowledge of the spin period. The contribution to $\dot{x}$ is only similar to
$\dot{x}_{\rm k}$ if the spin period is of the order of 0.5 ms, a rotational velocity
$\sim 3$ times faster than any pulsar observed to date. This is unlikely.

Not much changes if the companion is a WD: we still do not know its 
rotational angular momentum, $S$. The moment of inertia for a massive WD
is about $10^4$ times larger than for a NS. The shortest spin period known for a WD is
13.2 s \citep{2009Sci...325.1222M}. Thus, if the companion to NGC~1851A
were spinning at 13.2 s, the total angular momentum would be 
$S_{\rm p} \, \sim \, 4.8 \, \times \, 10^{41} \, \rm kg \, m^2 \, s^{-1}$, and
\begin{equation}
\dot{x}_{\rm LT} \, \sim \, -1.1 \, \times \, 10^{-14} \sin{\delta_{\rm C}} \sin{\Phi^0_{\rm C}} \rm \, lts\ s^{-1},
\end{equation}
which would be of the order of half of the estimated $\dot{x}_{\rm k}$.

Finally, if the companion is a WD, there will be a contribution of the
classical spin-orbit coupling to $\dot{x}$, caused by the rotationally-induced
oblateness of the companion. This is given by \cite{1998MNRAS.298..997W}:
\begin{equation}\label{eq:QM}
  \dot{x}_{\rm QM}\, = \, x\left(\frac{2\pi}{P_{\rm b}}\right) Q \cot\textit{i} 
    \sin\delta_{\rm c} \cos \delta_{\rm C} \sin \Phi^0_{\rm C}
\end{equation}
where:
\begin{equation}
  Q \, = \, \frac{k_2 R^2_{\rm C} \hat{\Omega}_{\rm C}^2}{a^2(1- e^2)^2} \quad {\rm with} \quad 
 \hat{\Omega}_{\rm C} \, \equiv \, \frac{\Omega_{\rm C}}{(G m_{\rm C} /R_{\rm C}^3)^{1/2}},
\end{equation}
where $m_{\rm C} = 2.4 \, \times \, 10^{30} \, \rm kg$ is the companion mass in kg, $R_{\rm C}$ is its radius
($\sim \, 3000 \, \rm km$), and $k_2$ is its
apsidal motion constant, which is a dimensionless measure of the oblateness
of the companion; for WDs this is of the order of 0.1 \citep{2017MNRAS.464.4349B}.
For a spin period of 13.2 s, we have $\hat{\Omega} \, \sim 0.2$, thus
$Q \, \sim \, 7.8 \, \times \, 10^{-10}$ and
\begin{equation}
\dot{x}_{\rm QM} \, \sim \, 8\, \times\, 10^{-14}  \sin\delta_{\rm c} \cos \delta_{\rm C} \sin \Phi^0_{\rm C}  \rm \, lts\ s^{-1},
\end{equation}
which, depending on the angles, could be few times larger than $\dot{x}_{\rm k}$ and is comparable in 
magnitude with $\dot{x}'$. 

Although this is unlikely, we cannot exclude the possibility of a fast-rotating WD companion.
If we have a large contribution of $\dot{x}_{\rm LT}$ and $\dot{x}_{\rm QM}$ to $\dot{x}$, we have no way
of separating it from the other effects.

\subsection{Influence of the proper motion on the Einstein delay}
\label{sec:pm_gamma}

As we have seen in the previous section, unless the companion is a 
WD with a very fast rotation, $\dot{x}_{k}$ is by far the
dominant contribution to $\dot{x}$, amounting up to $\pm$ 6.5 \% of $\dot{x}'$.
We will from now on assume that this is, indeed, the case.

Inverting equation~(\ref{eq:gamma_xdot}), and using equation~(\ref{eq:omdot_k}),
we obtain the variation of the actual $\gamma$ as a function of the proper
motion $\mu$, $\Omega$, $i$ and the measured $\gamma$ for systems like NGC~1851A:
\begin{equation}
\gamma(\Omega, i) \, = \, \gamma\, + \, \frac{x \sqrt{1 - e^2} } {\sin \omega} \frac{\mu}{\dot{\omega}} \cot i \sin(\Theta_{\mu} - \Omega),
\end{equation}
from this we obtain maximum and minimum values of $\gamma(\Omega,i)$
of 23.1 and 20.3 ms respectively (the ``measured'' value, $\gamma$, is 21.6 ms)
for values of $i$ close to the values derived in section~\ref{sec:gamma}.
These differences are slightly larger 
than the uncertainty of the measured $\gamma$, which is about 0.9 ms.
Using equation~(\ref{eq:gamma_masses}), we can then
obtain maximum and minimum companion masses of 1.263 and 1.144 $\rm M_{\odot}$;
again these differences (of the order of $0.06\, \rm M_{\odot}$) are larger
than the mass uncertainties derived in section~\ref{sec:gamma}
from the uncertainty of the measured $\gamma$,
$0.038\, \rm M_{\odot}$.

We note that these estimates rely on our current measurement
of the proper motion which, for the reasons discussed in section \ref{sec:proper_motion}) is not yet fully trustworthy.
If it is closer to the smaller GAIA proper
motion of NGC 1851, then we would also have a smaller
mass uncertainty caused by the proper motion.

\subsection{Shapiro delay}
\label{sec:shapiro}

The far from edge-on inclination means that the Shapiro delay is not 
easy to measure. In order to quantify its detectability
we use the orthometric parameterization of \citet{2010MNRAS.409..199F},
which is implemented as the DDFWHE model. To do this, we need first
a numerical value for the orthometric ratio $\varsigma$, this can be
derived from the $s\, \equiv\, \sin i$ estimated in section~\ref{sec:gamma}:
\begin{equation}
\varsigma \, = \, \frac{s}{1 + \sqrt{1 - s^2}} \simeq 0.50
\end{equation}
Fixing this in the model, we fit the othometric amplitude,
obtaining $h_3 \, = \, 0.2 \, \pm \, 1.4 \, \rm \mu s$.
This means that the Shapiro delay is not detectable in this system. However, this
value of $h_3$ is 1-$\sigma$ consistent with the expectation for this system,
$h_3\, = \, M_c T_{\odot} \varsigma^3 \, \simeq \, 0.74\, \rm \mu s$.

Despite being a non-detection, this constraint can already
exclude inclinations close to edge-on, as seen in figure~\ref{fig:masses}.

\begin{figure}
\centering
  \includegraphics[width=\columnwidth]{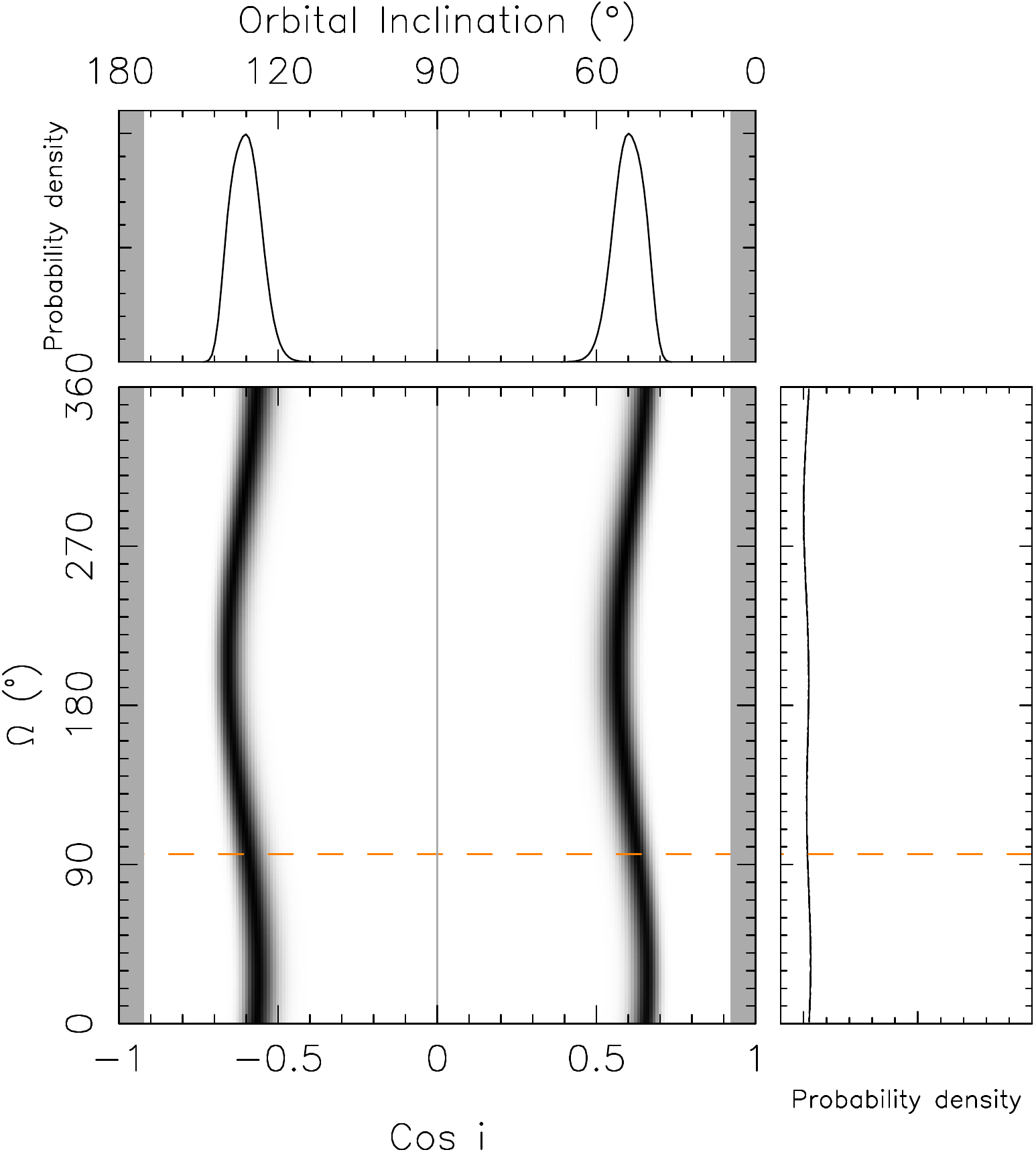}
  \caption[]{{\em Central panel}: the full $\cos i$-$\Omega$ space for binary pulsars.
  For PSR~J0514$-$4002A, the gray regions are
  excluded by the requirement that the pulsar mass must be larger than 0. The dotted orange line indicates
  the position angle of the proper motion of the system. The grey scale indicates the probability density,
  with zero indicated by white and maximum probability density indicated by black. As we can see, the
  best orbital inclination varies significantly with $\Omega$. {\em Top panel}: probability density
  function for $\cos i$, normalized to the maximum. {\em Right panel}: probability density function for
  $\Omega$. This is almost uniform, i.e., this variable is not constrained for this system.}
  \label{fig:om}
\end{figure}

\section{Bayesian analysis of the component masses}
\label{sec:masses}

In order to estimate the influence of the proper motion on $\gamma$,
we used in the previous section a value of $\cot i$ that is itself 
derived from the mass function of the system and the component masses.
The problem is that the component masses must be derived from $\gamma(\Omega, i)$
itself.

For this reason, we have implemented a Bayesian analysis of the system, with the aim of determining
the masses and orbital inclinations of the system in a fully self-consistent manner.

\subsection{Mapping the orbital orientation space}

In what follows, we roughly follow the Bayesian analysis done by
\cite{2018arXiv180905064S}, but with a few important differences.
As in the latter work, we map the quality of fit (the residual
$\chi^2$) for the orbital orientation space ($\cos i$ and $\Omega$) using the DDK
orbital solution. The full space ranges from $\cos i \, = \, -1$ to 1
and from $\Omega \, = \, 0\, \deg$ to $360 \, \deg$.
Randomly oriented orbits will populate this space uniformly.
In practice, we limit the range of $i$ to regions where $M_{\rm p}$ is positive,
i.e., where $\sin i \, > \, 0.3888$. Thus $-0.9213 \, < \, \cos i \, < \, 0.9213$.

Unlike \cite{2018arXiv180905064S}, we do not sample the
third dimension (in the latter case $M_{\rm tot}$) because, as discussed
in section~\ref{sec:omdot}, the changes in the total mass
caused by the kinematic contribution $\dot{\omega}_{\rm k}$ are of
the same order of the measurement uncertainty for $\dot{\omega}$. The
resulting changes in $M_{\rm tot}$ with $\Omega$ are two orders of magnitude
smaller than the uncertainties in the masses of the components; therefore
they are irrelevant for the estimates of the component masses. For
this reason, the whole map assumes one value of $M_{\rm tot}$ for the
estimation of the component masses.

For each point in the grid of $\cos i$ and $\Omega$ values, we introduce
these values in the DDK model via the \texttt{TEMPO}'s KIN and KOM parameters.
From this, the model internally estimates all kinematic effects, particularly
the secular $\dot{x}$. 

For each value of $i$, we derive $M_c$ from equation~(\ref{eq:sini}). We then introduce it in the model
via the \texttt{TEMPO}'s M2 parameter. This completes the description of the Shapiro delay:
its first parameter is the orbital inclination $i$.

Then, for each point we estimate $\gamma$ from $M_c$ and $M_{\rm tot}$ using
equation~(\ref{eq:gamma}), we introduce the result in the model via \texttt{TEMPO}'s GAMMA parameter.

All these parameters ($i$, $\Omega$, $M_c$ and $\gamma$) are fixed inputs to the
DDK model used to do the timing analysis for that grid point.
We then run {\tt TEMPO}, fitting for all other relevant timing parameters,
including $\dot{\omega}$, which is allowed to oscillate a little around the
best-fit model because of $\dot{\omega}_{\rm K}$, and $\dot{P}_{\rm b}$,
which is dominated by kinematic effects.
We then record the value of the $\chi^2$ for each combination of $\Omega$ and $\cos i$.
The resulting 2-D grid of $\chi^2$ values are then used to calculate a 2-dimensional
probability density function (pdf) for $\Omega, \cos i$, as discussed by
\cite{2002ApJ...581..509S}:
\begin{equation}
p(\Omega, \cos i) \propto e^{\frac{\chi^2_{\rm min} - \chi^2}{2}},
\end{equation}
where $\chi^2_{\rm min}$ is the lowest $\chi^2$ of the whole grid.
A greyscale plot of this pdf is displayed in the central plot of Fig.~\ref{fig:om}.

This 2-D pdf is then projected along two axes, $\cos i$ (1-D pdf is shown 
on top left panel in Fig.~\ref{fig:masses} and top panel in Fig.~\ref{fig:om})
and $\Omega$ (1-D pdf in right panel of Fig.~\ref{fig:om}). This is
then translated into the $M_c$ axis using
equation~(\ref{eq:sini}) (1-D pdf in right panel of Fig.~\ref{fig:masses}).
The 1-D pdf for $M_{\rm p}$ (top right panel of Fig.~\ref{fig:masses}) is merely a
reflection of the pdf for $M_c$.

\subsection{Results}

In Fig.~\ref{fig:om}, we can see how the orbital inclination
derived from $\dot{\omega}$ and $\gamma$ varies as a function of
$\Omega$, this is a visual demonstration of the effect of the
proper motion on $\gamma$. These orbital inclinations have
identical probabilities because the Shapiro delay is not measured
with enough precision to further restrict $\cos i$.
One of the consequences of this is that, as we can see on the
right plot, the probability density function for $\Omega$ is nearly constant.

The derived pulsar mass is $1.25^{+0.05}_{-0.06} \, M_{\odot}$
to 68.3 \% confidence limit (C.L.) and $1.25^{+0.09}_{-0.12} \, M_{\odot}$
to 95.4 \% C.L.; the asymmetry of the pdf can be easily be
seen in Fig.~\ref{fig:masses}. This mass is slightly lower, but
consistent, with the simple estimate made in section~\ref{sec:gamma}.
For the companion mass, the distribution is inverted: 
$M_c \, = \, 1.22^{+0.06}_{-0.05} \, M_{\odot}$ to 68.3 \% C.L.
and $1.22^{+0.12}_{-0.09} \, M_{\odot}$ to 95.4 \% C.L.

The mass of NGC~1851A is one of the lowest MSP masses measured to date.
There is only one other MSP, PSR~J1918$-$0642 ($M_{\rm p} \, = \, 1.29^{+0.10}_{-0.09} \, M_{\odot}$
\citealt{2018ApJS..235...37A}) that could have such a low mass.
This measurement demonstrates that the recycling to a spin period
of $\sim$5 ms can be achieved with a small amount of mass:
even if the system formed with the lowest known NS mass,
$1.174(4)\, M_{\odot}$ \citep{2015ApJ...812..143M}, the recycling process
would have been accomplished with a mass transfer of $\sim \, 0.08 \, M_{\odot}$.

On the other hand, the mass we measured for the companion implies that it can
be also a NS. We will explore this possibility in the following section.

\section{Companion Search}
\label{sec:companion_search}

If the companion is a NS, it could in principle also be a radio pulsar. For this reason, we carried out a search for pulsations from the putative companion.

For each observation listed in Table \ref{tab:NGC1851A_uGMRT_observations} (taking the CDP data whenever available, PA data otherwise) we first created a mask with the {\tt rfifind} routine of \PRESTO, in order to exclude all those frequency channels and time intervals affected by RFI. Taking the masks into account, we then used the {\tt prepdata} routine to de-disperse all the data at the nominal DM of NGC~1851A ($52.14$~\dmunit), scrunch the frequency band and create a {\em barycentered} (i.e. referred to the SSB) time series for each observation. In doing so, the CDP observations were also downsampled by a factor of eight, in order to match the sampling time of the PA data.

The actual search for the radio pulsations was done as follows. First, we used the {\tt PYSOLATOR}\footnote{\url{https://github.com/alex88ridolfi/pysolator}} software package to remove the orbital motion of the putative companion pulsar. In essence, {\tt PYSOLATOR} subtracts the predicted orbital R{\o}mer delay, together with any other relativistic effects, from each sample of the time series. As a result, it outputs a new {\rm demodulated} time series, where the pulsar appears as if it were isolated and located at the binary barycentre. By subtracting the orbital motion, the observed spin period of the possible companion pulsar will appear constant within each observation, as well as across multiple observations\footnote{This is strictly true only if we ignore the intrinsic spin-down of the pulsar. Given the small time span of our dataset, this is a safe assumption.}. This means that there is no need to perform an {\rm acceleration} search, and our sensitivity will not be limited by the length of the single observation \citep{Ransom2001}. 
All of this is possible only if we have good knowledge of the system's mass ratio, $q = \Mp / \Mc$. This is because we know all the characteristics of the companion's orbit very precisely, with the exception of its projected semi-major axis, $\xc$. The latter is related to the projected semi-major axis of the pulsar orbit, $\xp$, as $\xc = q\, \xp$. Given the uncertainties on $\Mp$ and $\Mc$, the mass ratio is in the range $q = 0.929 - 1.111$ to 68.3\% C.L., therefore it is a parameter of our search.

Therefore, we tried several values of $q$ and, for each of them, all the demodulated time series were Fourier transformed with the {\tt realfft} routine of \PRESTO. The so produced power spectra, after being de-reddened and normalized with \PRESTO's {\tt rednoise}, were summed together to produce a single stacked power spectrum relative to each considered trial $q$ value. In order to properly sum the spectra together, each demodulated time series was  padded by {\tt PYSOLATOR} so as to artificially obtain all time series with the same number of samples as that of the longest observation. To avoid sudden jumps in the time domain, which would translate into artifacts in the Fourier domain,  the added samples were all set to the average value of the last 10\% of the original time series.  This approach, which was also recently used by, e.g., \citet{Cadelano+2018}, allowed us to retain phase coherence within each observation and at the same time to have the resulting power spectra with homogeneous characteristics (i.e. Fourier bin size and frequency span), thus being straightforward to sum together.

Clearly, the range of $q$ values had to be explored with a sensible choice of the step size, $\Delta q$. The latter was chosen by imposing that, in the case of the best trial value and the fastest possible companion pulsar considered (i.e. spinning at 1000 Hz), the {\em maximum} drift of the observed spin frequency in the Fourier domain would be smaller than the size of one Fourier bin. For our dataset, this resulted in  $\Delta q = 0.00018$, corresponding to a total of 1011 trial values of $q$. 
The so obtained 1011 stacked spectra were then searched with \PRESTO's \texttt{accelsearch}, allowing no acceleration. All the candidates were then sifted, removing duplicates and excluding those with a significance of $\sigma < 5.0$. The candidates that survived the selection criteria were then folded using the original search-mode data, so as to retain full frequency information. This was done by producing, for each candidate, an ad-hoc ephemeris of the putative companion, containing the candidate spin frequency, as well as the orbital parameters as derived from the considered $q$ value. The resulting diagnostic plots were then inspected by eye.

None of the candidates could be ascribed to an astrophysical pulsar-like signal.









\section{Discussion and conclusions}
\label{sec:conclusions}

\subsection{On the nature of the companion}

The mass measurements for this system make it 
comparable to the highly eccentric PSR~J1807$-$2500B in NGC~6544 \citep{2012ApJ...745..109L},
where the companion, with a mass of $1.2064(20)\, M_{\odot}$, might also be a NS.
The companions to these systems could also be massive WDs.

There are systems in the Galactic disk with similar eccentricities, the 
double neutron star systems (DNSs). One of them, PSR~J1811$-$1736, even has an orbital period
and orbital eccentricity similar to NGC~1851A \citep{2007A&A...462..703C}.
However, it is unlikely that NGC~1851A formed like a DNS: all pulsars in DNSs have much longer spin periods than NGC~1851A (the shortest is that of PSR~J1946+2052, 16.7 ms, \citealt{2018ApJ...854L..22S}, PSR~J1811$-$1736 itself
has a spin period of 104 ms); this is a consequence of the fast evolution of the
massive companion, which results in a relatively short accretion episode
and therefore not much time for spin-up.

From this we conclude that systems like PSR~J1807$-$2500B and NGC~1851A formed in
a secondary exchange encounter, a likely event in the core of dense clusters
like NGC~1851 and NGC~6544. Such encounters happen (by definition) after the pulsar was recycled by
accretion of mass and angular momentum from a lighter companion, which can
last long enough to spin up the pulsar significantly. During the encounter, a massive
degenerate object came to such a close distance to the earlier binary system that
a chaotic interaction ensued. In this case, the most likely result is the
ejection of the lighter component of the binary and the formation of 
a new more compact and eccentric binary consisting of the pulsar and the 
massive degenerate intruder.

A consequence of the exchange interaction is that we cannot use stellar
evolution arguments to clarify the nature of the companion of NGC~1851A. 
Given the possibility that the latter is a NS, we 
have made a deep search for radio pulsations from that companion, as
described in section~\ref{sec:companion_search}. No pulsations were found.
This means that the question of the nature of the companion remains open:
indeed, the non-detection of the companion is not conclusive since many
NS companions to DNS systems are not detectable
as pulsars either; the same is true for the vast majority of NSs in our Galaxy and in GCs.

It is highly unlikely that the companion is a main sequence star.
Superior conjunction happens about 5 minutes after periastron, and the
separation between the pulsar and its companion in the plane of the sky is
scarcely more than one solar radius \citep{Freire+2007}; a main-sequence
companion would almost certainly produce eclipses near superior conjunction,
which are not observed.

\subsection{Is the companion losing mass?}

The apparent detection of an intrinsic increase of the orbital period
is intriguing. As calculated in section~\ref{sec:pbdot}, if this
is caused by mass loss from the companion, then its is losing 
mass at a rate that is about $10^{4}$ times larger than
the current mass loss rate for the Sun.

This is interesting because \cite{Freire+2007} presented
some evidence (based on the scintillation timescale of the pulsar, which seems
to be inversely proportional to the orbital velocity of the pulsar
around the centre of mass of the system) of some mass loss from the companion.
Another possibility is that the $\dot{P}_{\rm b}^{\rm int}$ has a tidal origin, with rotational
energy of the companion being transferred to the orbit (thus increasing the period).

In either case, a confirmation of the large $\dot{P}_{\rm b}^{\rm int}$ would
imply that the companion
is not a NS, because a NS companion would not likely lose mass
at any appreciable rate (owing to its extreme gravity) or have a tidal interaction
with the pulsar. We are thus left with the possibility of a massive WD companion. If this is losing
mass, it must be at its late stages of formation, where the last vestiges of
its envelope are still being ejected, otherwise no tidal effects are possible.
We find that such a hypothesis is unlikely:
any progenitors of 1.22-$M_{\odot}$ WDs should have long disappeared from 
the stellar population of NGC~1851. 
In any case, optical observations of the companion to NGC~1851A are strongly encouraged.

\subsection{Continued timing}

Apart from the issue of the $\dot{P}_{\rm b}$, there is another unsolved
issue remaining, the proper motion. This is
very different than the proper motion of the cluster and would suggest the pulsar
is on an escape path. The other is the large and unexpected $\dot{P}_b$
for this system. Both issues could arise from the 10-year gap in timing, where the
phase evolution of the pulsar has not been measured.
Continued timing should allow a better measurement of the higher spin frequency
derivatives, the proper motion and $\dot{P}_{\rm b}$, potentially allowing a full
reconstruction of the spin evolution during the 10-year gap in observations.
These measurements will certainly help clarify the issues raised by their current values.

Continued timing with the CDP mode will also improve the measurement of
$\gamma$; its uncertainty scales with $T^{-3/2}$, where $T$ is the timing baseline.
This will result in much thinner black lines in Fig.~\ref{fig:om}, i.e, a much
more restricted range of $\Omega$ and $\cos i$ where the system might exist.

Furthermore, a few long observing sessions around periastron in CDP mode will
certainly improve the constraints on the orthometric amplitude of the Shapiro 
delay, $h_3$. This could in principle allow a measurement of the masses that is
independent of $\gamma$ and any possible contributions to $\dot{x}$ that it might have.
Restricting the range of inclinations would imply (as we can see looking at Fig.~\ref{fig:om}) a restriction of the possible values of $\Omega$.

\subsection{Conclusions}

In this paper, we have presented the results for the recent timing of
the NGC~1851A binary pulsar with the uGMRT. Combining our ToAs with those
obtained with GBT 10 years before, we greatly improve the precision of the
measurement of $\dot{\omega}$, thus obtaining a far more precise estimate
of the total mass of the binary. We also measure, for the first
time, the proper motion of the system and the relativistic Einstein delay,
$\gamma$. This is the first time this has been done for a system with an orbital
period larger than 10 hours. The detection is helped by the sheer magnitude
of the effect, which is 4.5 times larger than any measurement of
$\gamma$ made to date in a binary pulsar. This is also the first time
that a measured $\gamma$ is larger (in this case four times larger) than the spin
period of the pulsar.

The latter effect allows a measurement of the individual masses of the
components. One of the most important results in this paper is a detailed
study of the conditions under which $\gamma$ can be measured and its
covariance with the variation of the projected semi-major axis, $\dot{x}$,
in particular with the kinematic component of that term that arises
inevitably from the proper motion of the system.
This means that, in order to estimate the component masses of a wide
binary system with the help of $\gamma$, we must take into account at least
the effect of the proper motion. We do this in an economical and
self-consistent way by sampling the quality ($\chi^2$) of the timing fit for 
the full orbital orientation space (which consists of $\cos i$ and $\Omega$).
From this $\chi^2$ map, we derive probabilistic distributions for
$\cos i$, $M_{\rm p}$ and $M_c$. The median and 68.3\% confidence limits
of the mass distributions are given by
$M_{p} \, = \, 1.25^{+0.05}_{-0.06} \, M_{\odot}$
and a companion mass of $M_{c} \, = \, 1.22^{+0.06}_{-0.05} \, M_{\odot}$. 

The low mass of the MSP implies that the recycling process can be
achieved with a relatively small amount of mass, $< \, 0.08 \, M_{\odot}$.
We cannot use this number to estimate the efficiency of the recycling process
(as done by, e.g., \citealt{2012MNRAS.423.3316A}) because we do not know the
mass and orbital properties of the original donor star.

The mass of the current companion implies the possibility that it is
also a NS.
This makes the system very similar to PSR~J1807$-$2500B, located in
the globular cluster NGC~6544. Both systems were very likely formed
by exchange interactions in the core of the globular clusters where
they are located, both are potential MSP - NS systems, but each could
also be a MSP - massive WD system. Given the possibility that the companion
is a NS, we have looked deeply for radio pulsations from the
companion, but none were found. Therefore, we cannot determine the
nature of the companion with any certainty.

The measured masses imply, according to GR, that the time until gravitational-wave
induced merger of $\sim \, 463 \, \rm Gyr$, which is more than 30 times the Hubble time.
It is to be expected, given the dense environment and the history of the system,
that its interactions with other stars in the cluster will produce
very significant changes in its orbital parameters (or even in the companion itself)
on a timescale much shorter than the orbital decay timescale.

Future observations will refine the proper motion, which will allow
us to measure precisely (and hopefully accurately) the velocity
difference relative to the cluster. Such observations will also allow
a confirmation (or not) of the anomalous orbital period derivative
of the system. 

\section*{Acknowledgements}

A.R. and P.C.C.F. gratefully acknowledge financial support by the European Research Council,
under the European Union's Seventh Framework Programme (FP/2007-2013) grant agreement
279702 (BEACON) and continuing support from the Max Planck Society. A.R. thanks the Autonomous Region of Sardinia (RAS) for financial support through the Regional Law 7 August 2007 n. 7 (year 2015) ``Highly qualified human capital'', in the context of the research project CRP 18 ``General relativity tests with the Sardinia Radio Telescope'' (P.I.: M. Burgay). We thank the staff of the GMRT for help with the observations. The GMRT is operated by the National Centre for Radio Astrophysics (NCRA) of the Tata Institute of Fundamental Research (TIFR), India. The National Radio Astronomy Observatory is a facility of the National Science Foundation operated under cooperative agreement by Associated Universities, Inc. S.M.R. is a CIFAR Senior Fellow and is supported by the NSF Physics Frontiers Center award 1430284. A.R. also thanks Caterina Tiburzi, Andrea Possenti, Cees Bassa and Kuo Liu for useful discussions. This research has made extensive use of NASA's Astrophysics Data System (ADS).





%
%
%
%


\bsp	
\label{lastpage}

\end{document}